\documentclass[sigconf]{acmart}
\renewcommand\footnotetextcopyrightpermission[1]{}
\newcommand{\SysName}{\texttt{PromptTuner}}

\usepackage{balance}
\usepackage[normalem]{ulem}
\usepackage{graphicx}
\usepackage{algorithm}
\usepackage{algpseudocode}
\usepackage{amsthm}
\usepackage{amsmath}
\usepackage{multirow}
\usepackage{url}
\usepackage{subcaption}
\usepackage{enumitem}
\usepackage{makecell}
\usepackage{booktabs} 
\usepackage{soul}
\usepackage{textcomp}
\usepackage{pifont}
\usepackage{adjustbox}
\usepackage{wrapfig}

\usepackage{amssymb}
\usepackage{bbm}

\usepackage[normalem]{ulem}

\usepackage[most]{tcolorbox}

\newcommand{\cmark}{\ding{51}}%
\newcommand{\xmark}{\ding{55}}%

\usepackage{tikz}
\newcommand\encircle[1]{\tikz[baseline=(X.base)] 
    \node (X) [draw, shape=circle, inner sep=0pt, fill=black, text=white] {\strut #1};}

\theoremstyle{definition}

\newenvironment{denseitemize}{
	\begin{itemize}[topsep=2pt, partopsep=0pt, leftmargin=1.5em,label=$\star$]
		\setlength{\itemsep}{3pt}
		\setlength{\parskip}{0pt}
		\setlength{\parsep}{0pt}
	}{\end{itemize}}

\begin{document}


\title{\SysName: SLO-Aware Elastic System for LLM Prompt Tuning}

\author{Wei Gao$^{1}$, Peng Sun$^2$, Dmitrii Ustiugov$^1$,  Tianwei Zhang$^1$, Yonggang Wen$^1$}
\affiliation{%
  \institution{$^1$College of Computing and Data Science, Nanyang Technological University, \quad $^2$Unaffiliated}
  \country{gaow0007@e.ntu.edu.sg, sunpeng1@outlook.com, \{dmitrii.ustiugov,tianwei.zhang,ygwen\}@ntu.edu.sg}}

\renewcommand{\shortauthors}{Wei Gao, Zhisheng Ye, Peng Sun, Yonggang Wen, and Tianwei Zhang}

\renewcommand{\authors}{Wei Gao, Zhisheng Ye, Peng Sun, Yonggang Wen, and Tianwei Zhang}

\begin{abstract}
Prompt tuning has become a prominent strategy for enhancing the performance of Large Language Models (LLMs) on downstream tasks. Many IT enterprises now offer Prompt-Tuning-as-a-Service to fulfill the growing demand for prompt tuning LLMs on downstream tasks. Their primary objective is to satisfy users' Service Level Objectives (SLOs) while reducing resource provisioning costs. Nevertheless, our characterization analysis for existing deep learning resource management systems reveals that they are insufficient to optimize these objectives for LLM prompt tuning workloads.

In this paper, we introduce \SysName{}, an SLO-aware elastic system to optimize LLM prompt tuning. It contains two innovations. (1) We design a \textit{Prompt Bank} to identify efficient initial prompts to expedite the convergence of prompt tuning. (2) We develop a \textit{Workload Scheduler} to enable fast resource allocation to reduce the SLO violation and resource costs. In our evaluation, \SysName{} reduces SLO violations by 4.0\(\times\) and 7.9\(\times\), and lowers costs by 1.6\(\times\) and 4.5\(\times\), compared to INFless and ElasticFlow respectively. 
\end{abstract}

\maketitle

\section{Introduction}
\label{sec:introduction}

Large Language Models (LLMs) are becoming prevalent in many scenarios owing to their exceptional capabilities~\cite{claude_ai,google_gemini,openai2023gpt4}. LLM developers employ a compelling and widely embraced technique known as \textit{prompt tuning}, to customize LLMs for diverse applications, particularly  agentic ones, without altering the model weights~\cite{chen2023autoagents,midjourney,chatgpt,manus_website}. However, the manual process of prompt tuning is time-consuming and resource-intensive~\cite{do2024automatic,zhou2022large,wang2025sequential}, driving many IT companies to offer Prompt-Tuning-as-a-Service to enable automatic prompt tuning within seconds to minutes~\cite{prompthero,promptperfect,portkeyai}. In this business model, users furnish initial prompts and downstream datasets and select the base LLMs. Subsequently, the service provider must efficiently allocate GPUs to optimize the prompts for the given datasets, handling tens of thousands of LLM prompt tuning (LPT) requests per day~\cite{portkeyai}, returning the finalized prompts to users.



The service provider has several considerations when serving users' LPT requests. First, users concentrate on the accuracy\footnote{We use accuracy as a universal term to denote any evaluation metric.} and the latency of their LPT requests. They will specify the Service Level Objectives\footnote{The definition of SLO is explained in \S\ref{subsec:system-system-overview}.} (SLOs) of the targeted accuracy and latency~\cite{prompthero,promptperfect,mergeflow,portkeyai}. Second, the service provider rents top-grade GPU resources from clouds~\cite{alibabacloud,AWS,Azure} to handle users' LPT requests. Given the increasing number of LPT requests and the considerable cost of renting GPUs, there is a pressing need to design systems that optimize resource allocations for LPT workloads. Such optimization aims to reduce resource costs for service providers while fulfilling SLOs for users.




We present a workload characterization summary of LPT workloads in \S\ref{sec:characterization-prompt-tuning}, and find that they exhibit training-like and inference-like features. A straightforward approach is to leverage prior studies in cluster management systems for training and inference workloads to address LPT demands. However, our empirical study in \S\ref{sec:limitations-of-existing-solutions} shows that they are ineffective in managing LPT workloads. First, previous SLO-aware systems for Deep Learning (DL) training~\cite{Chronus,elasticflow,GENIE} oversubscribe a \textbf{fixed}-sized GPU cluster to guarantee SLOs, resulting in increased resource costs. Also, the commonly adopted frequent resource allocation could incur nearly one-minute resource allocation overhead for LLMs~\cite{thorpe2023bamboo,jang2023oobleck} and pose a significant barrier to enforcing minutes-level latency SLOs for LPT workloads. Second, prior inference systems~\cite{INFaaS, INFless,FaaSwap,zhang2019mark} adopt two techniques: (1) they autoscale the quantity of GPUs needed to reduce resource costs; (2) they pre-load the DL runtime (e.g., CUDA/framework runtime) and model weights in the GPU memory for a time period to reduce the allocation overhead and optimize the SLO attainment. However, these solutions adhere to a \textbf{fixed} GPU allocation, normally assigning one GPU for each job, compromising the adaptability required to meet varying levels of SLOs for LPT jobs. As shown in \S\ref{sub:dl-inference-limitations}, even with the incorporation of multi-GPU allocation into DL inference systems, they still struggle to serve LPT jobs effectively. Overall, prior training and inference systems exhibit deficiencies in realizing SLO satisfaction and cost reduction simultaneously for LPT.

Additionally, a unique feature of LPT workloads is overlooked by existing DL cluster management systems and LPT services: their model convergence is highly \textit{sensitive to the initial prompts} (\S\ref{sec:characterization-prompt-tuning}). This sensitivity suggests the significant variance in the number of iterations required to achieve the targeted accuracy given different initial prompts. For example, a well-curated initial prompt demands fewer tuning iterations than a poor one, thereby mitigating SLO violations and reducing resource costs. Practically, LLM developers adopt two initialization methods. First, the current practice of LPT services is manual initialization. Users are asked to craft initial prompts by themselves~\cite{yao2023tree,besta2023graph}. Alternatively, users are recommended to reuse publicly available prompts directly~\cite{promptbase,prompthero}. However, both practices rely on human expertise, substantial GPU resources, and time for these laborious trial-and-error processes. Second, some LLM studies~\cite{ye2023prompt,zhou2022large} and LLM services~\cite{claude} propose induction initialization to guide LLMs to automatically generate initial prompts without human expertise. However, the quality of the generated initial prompt heavily relies on the performance of the LLM itself~\cite{ye2023prompt,zhou2022large} (evaluated in \S\ref{sec:eval-ablation-studies}). Despite the potential benefits, few systematic efforts exist to automatically and efficiently identify initial prompts for a given LPT job.

To bridge these gaps, we design \SysName{}, an SLO-aware elastic cluster management system dedicated to LPT. \SysName{} consists of two designs. First, we design a \textit{Prompt Bank} as a query engine to automatically and efficiently search the initial prompt for a given LPT job. The observation that prompts optimized for one LPT task can serve as effective initial prompts for another task with high similarity motivates the design of the Prompt Bank~\cite{su-etal-2022-transferability,NLPTransfer}. As public prompts optimized for various tasks are noticeably increasing~\cite{awesome_chatgpt_prompts}, we collect thousands of high-quality prompts as the initial prompt candidates for incoming LPT jobs. We adopt a two-layer data structure that enables quick search of an effective initial prompt, reducing the selection time to under 10 seconds.




Second, we design a \textit{Workload Scheduler} that supports fast and elastic GPU allocation for LPT workloads to meet SLOs and reduce resource costs. The Workload Scheduler allows LPT jobs based on the same LLM to reuse the GPUs from a warm GPU pool comprising GPUs with the same job-specific pre-loaded LLM runtime and weights, providing rapid GPU allocation. The Workload Scheduler maintains a dedicated warm GPU pool for each LLM and dynamically adjusts the pool size by adding GPUs from a shared cold GPU pool. It consists of two algorithms to manage these GPU pools. The first delivers fast multi-GPU allocation from the warm GPU pools to LPT jobs, reducing the considerable initialization overhead for multi-GPU execution (\S\ref{sub:dl-inference-limitations}). The second one dynamically adjusts the number of GPUs for each warm GPU pool to balance the trade-off between SLO attainment and resource costs in the dynamic traffic of LPT jobs. Furthermore, the Workload Scheduler intelligently decides whether to route incoming LPT requests to the Prompt Bank or execute them directly to prevent resource contention. 




We implement \SysName{} atop Knative, and select three LLMs (GPT2-Base~\cite{GPT-3}, GPT2-Large~\cite{GPT-3}, Vicuna-7B~\cite{vicuna}) to compare \SysName{} with the state-of-the-art SLO-aware DL inference system INFless~\cite{INFless} and training system ElasticFlow~\cite{elasticflow} on a cluster of 32 NVIDIA A100-80GB GPUs. \SysName{} reduces the SLO violation by up to 4.0\(\times\) (INFless) and 7.9\(\times\) (ElasticFlow), and reduces the cost by up to 1.6\(\times\) (INFless) and 4.5\(\times\) (ElasticFlow). We also conduct experiments on LLaMA-30B and DeepSeek-R1-Distill-Qwen-7B (Qwen7B-R1) and perform large-scale experiments in a 96-GPU cluster to demonstrate its superiority under heavy workload settings. Our contributions are as follows:


\begin{denseitemize}

\item We perform an in-depth characterization analysis for LPT workloads and conduct detailed empirical studies to uncover the inefficiencies of existing systems to handle LPT. 

\item We present \SysName{}, an elastic system for LPT workloads that can guarantee SLOs for users and reduce resource costs for service providers. 

\item We perform extensive evaluations to validate the efficiency of the \emph{Prompt Bank} and the \emph{Workload Scheduler}. 




\end{denseitemize}

\section{LPT Workload Characterization}
\label{sec:workload-characterization}


\begin{figure}[t]
\centering
  \centering
  \includegraphics[page=1, width=0.49\textwidth]{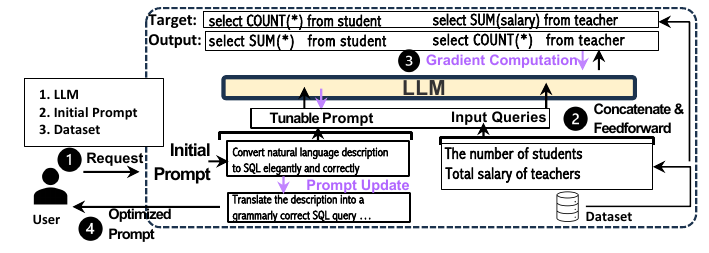}
\vspace{-15pt}
    \caption{\small An example of LLM prompt tuning. The user first prepares the LLM, the initial prompt, and the task-specific dataset, which consists of a batch of input queries and target responses. During the execution stage, it optimizes the tunable prompt starting from the initial prompt on the given dataset.}
\vspace{-15pt}
\label{fig:illustration-pt}
\end{figure}

\subsection{Prompt Tuning}
\label{sec:motivation-prompt-tuning-in-gpu-datacenters} 
Prompt tuning is an approach to obtain high-quality responses for a specific task from an LLM by attaching a prompt prefix (simply referred to as prompt), saving the high cost of retraining the model weights. An LPT job optimizes a prompt that elicits the best response from the LLM when prepended to an input query. Figure~\ref{fig:illustration-pt} shows an example of the task of converting the natural language to SQL language using the gradient-based LPT algorithm~\cite{li2021prefix}. The user sends an LPT request containing the LLM, the initial prompt ("\textit{Convert natural language description to SQL elegantly and correctly}"), and the task-specific dataset consisting of input queries and corresponding target responses. Some LPT service providers~\cite{promptbase,prompthero} recommend the user to specify their initial prompt based on their expertise ({\tiny\encircle{\normalsize1}}). To execute an LPT job, the LPT system feeds this set of input queries into the LLM ({\tiny\encircle{\normalsize2}} ). The system runs the given LPT algorithm to compute the loss between the generated output sentences and targeted responses. Then it backpropagates the textual gradients and updates them on the tunable prompt ({\tiny\encircle{\normalsize3}}). After multiple iterations, the optimized prompt is generated: "\textit{Translate the description into a grammatically correct SQL query optimized for speed and accuracy}", and returned to the user ({\tiny\encircle{\normalsize4}}). 


\noindent\textbf{Prevalence of LPT Workloads.} Today, LPT workloads emerge as an important GPU consumer, making prompt-tuning services an essential business practice~\cite{prompthero}. When a user sends an LPT request, the system registers it as an LPT job and schedules each LPT job to run on GPUs while maintaining the strict SLOs the users impose. 

The prevalence of LPT workloads manifests in three aspects.  First, many prompt-tuning services~\cite{promptperfect,prompthero,promptbase,promptrr,flowgpt,textcortex,portkeyai} serve to expand LLMs across various fields, making the LLM prompt market trendy and growing. LLM developers daily produce tens of thousands of prompt-tuning requests~\cite{prompthero,promptperfect,portkeyai} and claim a significant number of high-grade GPUs~\cite{nvidia2020a100whitepaper,nvidia2022h100whitepaper} to respond to these LPT requests quickly. Second, LLM developers utilize curated prompts to guide commercial LLM services in surpassing human-engineered prompts on downstream tasks~\cite{zhou2022large}. As shown in the second and third columns of Table~\ref{tab:llm-benefits}, prompt tuning surpasses few-shot prompting techniques across 10 LLM tasks~\cite{honovich2022instruction}, delivering an average improvement of 2.5\(\times\) and 1.8\(\times\) with GPT-3.5 and GPT-4, respectively. Third, LLM developers rely on prompt tuning methods to enhance the accuracy of open-source LLMs on specific tasks~\cite{PoT,InterVenor,RelIndex}. In the forth, fifth and sixth columns of Table~\ref{tab:llm-benefits}, prompt tuning achieves an average score improvement of 5.4\(\times\), 4.0\(\times\), 2.2\(\times\) across various tasks on Vicuna-7B, LLaMA-30B, and Qwen7B-R1, respectively. Open-source LLMs provide access the output at the logits layer, leading to higher accuracy compared to commercial LLM services.

\begin{table}[]
\caption{\small The average score ($\uparrow$) of prompting techniques over tasks.}\label{tab:llm-benefits}%
\vspace{-8pt}
\centering
\resizebox{0.95\linewidth}{!}{
\begin{tabular}{lrrrrrr}\toprule
\textbf{Prompting Techniques} &\textbf{GPT-3.5} &\textbf{GPT-4} &\textbf{Vicuna-7B} &\textbf{LLaMA-30B } & \textbf{Qwen7B-R1}\\\midrule
Few Shot &29.8  & 37.0 &14.9 &20.8 & 34.1 \\
Prompt Tuning &70.1  &75.8 &80.5 &84.1 & 85.7 \\
\bottomrule
\end{tabular}}
\vspace{-10pt}
\end{table}

\begin{figure}[t]
\centering
\begin{subfigure}{.32\linewidth}
  \centering
  \includegraphics[width=\linewidth]{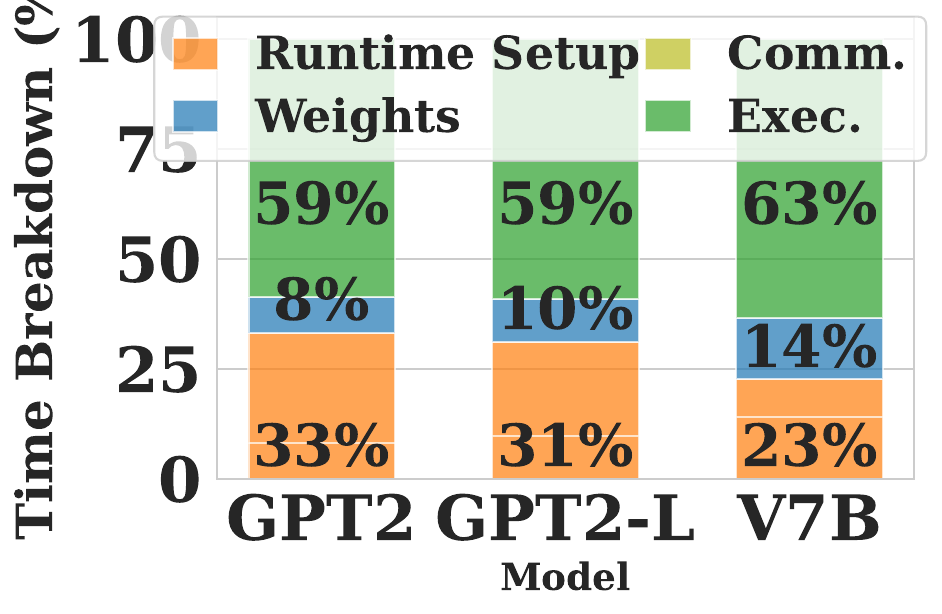}
  \caption{Time Breakdown.}
  \label{fig:profile-weight-ratio}
\end{subfigure}
\begin{subfigure}{.32\linewidth}
  \centering
  \includegraphics[width=\linewidth]{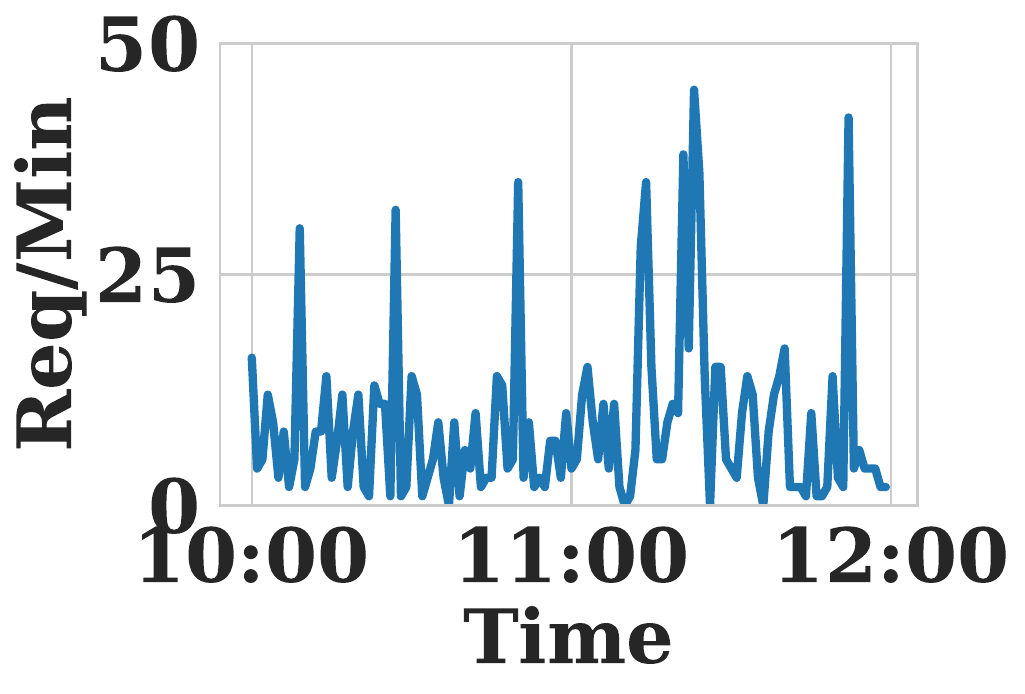}
  \caption{Trace Pattern.}
  \label{fig:trace-pattern}
\end{subfigure}
\begin{subfigure}{.32\linewidth}
  \centering
  \includegraphics[width=\linewidth]{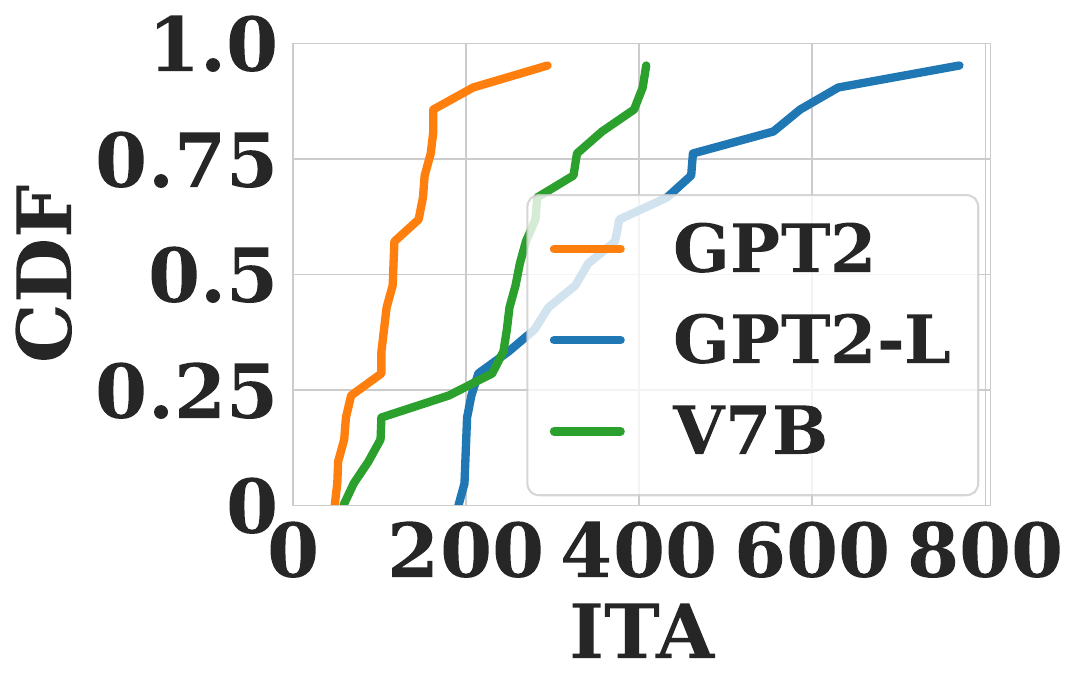}
  \caption{ITA CDF.}
\label{fig:profile-dist}
\end{subfigure}
\vspace{-10pt}
\caption{Characteristics of LPT workloads: (a) The end-to-end LPT job execution time breakdown across different LLMs. (b) A 2-hour LPT workload trace from a cluster. (c) The Iteration-To-Accuracy (ITA) distribution of various initial prompts with the \texttt{SAMSUM} dataset~\cite{SAMSUM} across different LLMs.}
\vspace{-15pt}
\label{fig:motivation-lpt-characterization-inference}
\end{figure}

\subsection{Prompt Tuning Workload Characteristics}
\label{sec:characterization-prompt-tuning}
Next, we study the LPT workload characteristics, which can guide the design of an efficient LPT cluster management system. We experiment with three popular LLMs (GPT2-Base, GPT2-Large, Vicuna-7B) and the \texttt{SAMSUM} dataset~\cite{SAMSUM} on a server of 8 Nvidia A100-80GB GPUs. We identify some common characteristics that LPT workloads share with training and inference workloads, which have been extensively studied by prior works~\cite{ServerlessATC20,INFaaS,FaaSwap,Philly,Helios,MLaaS}.

\noindent\textbf{Synchronous Cross-GPU Communication.}
Similar to DL training workloads, executing an LPT job requires iterations of feed-forward/backward passes, followed by a synchronous exchange of prompt gradients after each iteration. However, the cross-GPU communication of LPT is much lower than that of DL training. Figure~\ref{fig:profile-weight-ratio} shows the time breakdown of three LPT workloads: the communication overheads are within 0.4-0.5\% of the total execution time. Hence, LPT workloads can enjoy a nearly linear throughput increase when the number of allocated GPUs is increased. 


\noindent\textbf{Dynamic Traffic.} 
LPT is a user-facing service, featuring highly volatile dynamic traffic. We analyze a trace of LPT jobs sampled from a 64-GPU cluster in a large institute (anonymized). Figure~\ref{fig:trace-pattern} presents the LPT job arrival time for prompt-tuning Vicuna-7B within two hours. We observe large spikes of LPT traffic, with the maximum number of requests per minute being 5\(\times\) the mean. Such a pattern indicates that an efficient LPT system needs highly reactive autoscaling.



\noindent\textbf{High GPU Allocation-to-Computation Ratio.} The dynamic nature of LPT workloads requires fast provisioning of GPUs, similar to inference workloads~\cite{ServerlessATC20,INFaaS,FaaSwap}. We measure the GPU allocation overhead (including container setup, framework initialization, and GPU runtime creation), which accounts for 37-41\% of the total execution time. This indicates the need for fast GPU provisioning and reuse across LPT jobs.

\noindent\textbf{High Sensitivity to Initial Prompts.} We observe that the convergence speed of the LPT workload highly depends on the choice of the initial prompt. We measure the convergence speed with the Iterations-To-Accuracy (ITA) metric using 20 randomly selected prompts on the \texttt{SAMSUM} dataset~\cite{SAMSUM} for different LLMs. Figure~\ref{fig:profile-dist} shows the cumulative distribution function (CDF) of the ITA metric. The median and maximum ITA values are 1.7-4.5\(\times\) higher than the minimum ITA, indicating the significance of selecting an effective initial prompt at the beginning of LPT. Given the availability of substantial public prompts~\cite{awesome_chatgpt_prompts}, we identify the possibility of finding and reusing them as initial prompts for specific tasks. 

\noindent\textbf{Characterization Summary.} Table~\ref{tab:workload-comparison} summarizes the characteristics of LPT, DL training, and inference workloads. First, LPT workloads require synchronous communication after each iteration, similar to DL training. Second, LPT workloads are highly dynamic and suffer from lengthy GPU allocation delays, similar to DL inference workloads. Meanwhile, LPT workloads have a unique feature: their processing time highly depends on the choice of the initial prompts.





\begin{table}[]
\caption{\small Comparison of LPT, DL inference and training workloads.}\label{tab:workload-comparison}%
\vspace{-5pt}
\centering
\resizebox{0.8\linewidth}{!}{
\begin{tabular}{lccc}\toprule
\textbf{Characteristics}     & \textbf{LPT} & \textbf{Inference} & \textbf{Training}     \\ \midrule
Synchronous Cross-GPU Comm. & \cmark    & \xmark       & \cmark \\ \midrule
Dynamic Traffic                           & \cmark       & \cmark   & \xmark       \\ \midrule
High Allocation Overhead    & \cmark          & \cmark      & \xmark          \\ \midrule
Prompt Sensitivity          & \cmark          & \xmark       & \xmark          \\ \bottomrule
\vspace{-30pt}
\end{tabular}}
\end{table}

\section{Characterization of the Existing Cluster Management Systems}
\label{sec:limitations-of-existing-solutions}

As LPT workloads share similar execution features with DL training and inference workloads, a natural strategy is to extend existing cluster management systems for training and inference to LPT. In this section, we quantitatively evaluate the efficiency of state-of-the-art inference and training systems using the same experimental setup as in \S\ref{sec:experimental-setup}. We use the first 20 minutes of the trace in Figure~\ref{fig:trace-pattern} to run the prompt-tuning jobs based on the Vicuna-7B model. 





\subsection{Inefficiency of DL Training Systems}
\label{sub:dl-training-limitations}

Prior works have proposed many system designs ~\cite{Chronus,UniSched,elasticflow,GENIE,yang2023hydra} that optimize the execution of DL training workloads. These systems provision a fixed-size GPU cluster, further referred to as a GPU pool, and frequently allocate GPUs from this pool to jobs to maximize GPU utilization. 


We evaluate the efficiency of the state-of-the-art SLO-aware training system ElasticFlow~\cite{elasticflow}. It dynamically adjusts the number of allocated GPUs for each job to improve the job throughput and SLO attainment. However, in ElasticFlow, the resource costs per time unit remain fixed for all statically provisioned GPUs, regardless of actual usage. Figure~\ref{fig:gpu-usage} shows the GPU cluster utilization of ElasticFlow. On average, ElasticFlow only achieves 56\% GPU cluster utilization, almost doubling the GPU cluster's cost.




\vspace{-5pt}
\begin{tcolorbox}[colback=gray!5!white,colframe=gray!75!black,left=1mm, right=1mm, top=0.5mm, bottom=0.5mm, arc=1mm]
    \textbf{Inefficiency 1}: The static provisioning of a fixed-size GPU cluster in existing DL training systems results in a high resource cost when running LPT workloads.
\end{tcolorbox}


\subsection{Inefficiency of DL Inference Systems}
\label{sub:dl-inference-limitations}
Existing inference systems~\cite{INFaaS, INFless,FaaSwap,zhang2019mark} often feature a serverless autoscaling architecture. Upon receiving an inference job, they typically allocate a GPU-equipped container, also called an instance that is a unit of scaling, from a large pool of available GPUs to the provider for each incoming job. To alleviate the lengthy GPU container startup overheads, providers keep idle instances ready to serve any future inference jobs of the same model, occupying pricey GPU memory for a prolonged time. These systems implement autoscaling to adjust the number of instances for each model according to changes in the inference traffic.

Although these designs avoid the static resource provisioning of the training systems, their performance suffers from other limitations. First, they scale the resources of each model independently without considering a globally optimal schedule. Second, prior systems~\cite{INFaaS, INFless,FaaSwap,zhang2019mark} are limited to allocating one GPU for each instance of a model. Last, they lack support for synchronous cross-GPU communication, which is required for LPT jobs. 



\begin{figure}[t]
\centering
\begin{subfigure}{.32\linewidth}
  \centering
  \includegraphics[width=\linewidth]{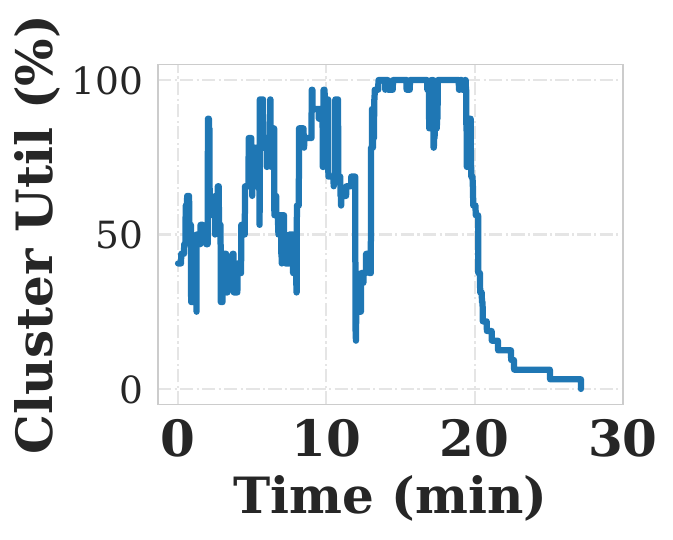}
  \caption{ElasticFlow.}
  \label{fig:gpu-usage}
\end{subfigure}
\begin{subfigure}{.32\linewidth}
  \centering
  \includegraphics[width=\linewidth]{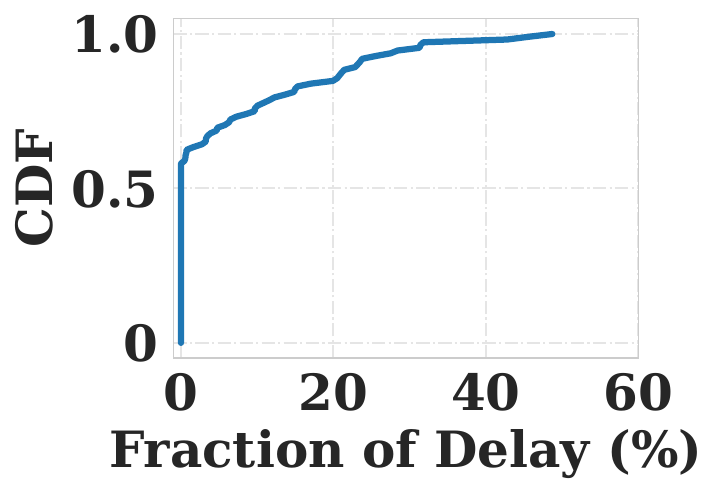}
  \caption{INFless.}
  \label{fig:inference-cost-dmr}
\end{subfigure}
\begin{subfigure}{.32\linewidth}
  \centering
  \includegraphics[width=\linewidth]{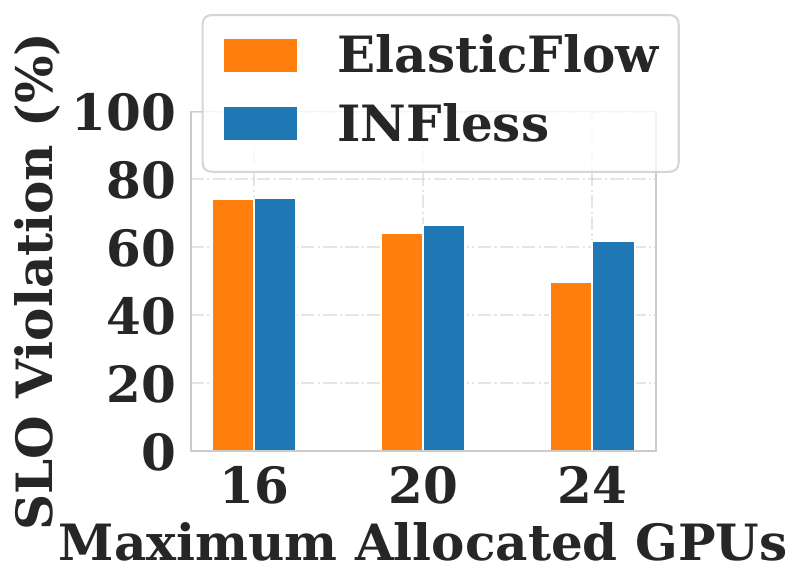}
  \caption{SLO violations.}
\label{fig:both-slo-violation}
\end{subfigure}
\vspace{-10pt}
\caption{\small Characterization of existing DL systems: (a) The cluster utilization (\%) ($y$-axis) in ElasticFlow over time ($x$-axis). (b) The CDF ($y$-axis) illustrates the fraction ($x$-axis) of waiting delay in the end-to-end latency caused by the instance initialization.(c) SLO violation (\%) of ElasticFlow and INFless across varying maximum GPUs.}
\vspace{-15pt}
\label{fig:empirical-studies}
\end{figure}

We select INFless~\cite{INFless}, a representative DL inference system, for our evaluation. However, running an LPT job on a single instance is insufficient to improve the job throughput and meet the emergent latency SLOs. To address this limitation, we extend INFless to support synchronous cross-GPU communication via Memcached \cite{Memcached}, as commonly used in serverless systems~\cite{SIREN,lambdaml}. The implementation details of multi-GPU execution can be found in~\S\ref{sec:preemptive-elastic-execution}. Thus, a single LPT job can use multiple instances, i.e., multiple GPUs, to accelerate its completion. Nonetheless, in INFless and other inference systems, some instances may need tens of seconds to initialize, thereby incurring long waiting time for the LPT job when running across multiple instances. Figure \ref{fig:inference-cost-dmr} depicts that instance initialization contributes on average 11\% to the end-to-end LPT job latency, and up to 50\% in the worst case.



\begin{tcolorbox}[colback=gray!5!white,colframe=gray!75!black,left=1mm, right=1mm, top=0.5mm, bottom=0.5mm, arc=1mm]
    \textbf{Inefficiency 2}: 
    The initialization techniques for a single instance adopted by DL inference systems incur substantial delays for the initialization of multi-GPU instances. 
\end{tcolorbox}



\vspace{2pt}
Unsurprisingly, ElasticFlow and INFless show substantially high SLO violation rates due to the abovementioned inefficiencies. Figure~\ref{fig:both-slo-violation} shows the SLO violation (\%) -- up to 70\% -- occurring when executing the LPT workload on top of ElasticFlow and INFless with varying maximum numbers of allocated GPUs. These results demonstrate that existing cluster management systems are unsuitable for LPT workloads, calling for designing a new system tailored to the LPT workload characteristics in \S\ref{sec:characterization-prompt-tuning}.

\section{System Design}
We introduce \SysName{}, an SLO-aware elastic cluster management system for LPT workloads. We begin  with the design insights and overview, followed by the illustration of its two key components: Prompt Bank and Workload Scheduler.

\subsection{Design Insights}
\label{sec:design-insight}


The design of \SysName{} is motivated by two insights. Our first insight is that \emph{LPT tasks can reuse the prompts optimized for similar tasks as their initial prompt to reduce the number of tuning iterations needed to achieve the desired accuracy.} Extensive empirical studies on transfer learning~\cite{NLPTransfer,su-etal-2022-transferability} and theoretical analysis~\cite{tripuraneni2020theory} affirm that reusing prompts can considerably accelerate the model convergence. However, the key to successfully reusing prompts lies in automatically and promptly identifying the most effective ones.

Our second insight is that \emph{LPT workloads can reuse the runtime of the jobs based on the same LLMs}. Indeed, many LPT jobs load the same runtime state into the GPUs before execution~\cite{S-LORA,chen2023punica}. This state includes the CUDA and DL framework dependencies and model weights. Reusing this state can substantially reduce the GPU allocation overhead (\S\ref{sec:characterization-prompt-tuning}). However, the key to effectively reusing the runtime lies in mitigating the substantial delays for LPT jobs demanding multiple GPUs, as emphasized in \S\ref{sub:dl-inference-limitations}.

\begin{figure}
\centering
  \centering
  \vspace{-10pt}
  \includegraphics[page=1, width=0.45\textwidth]{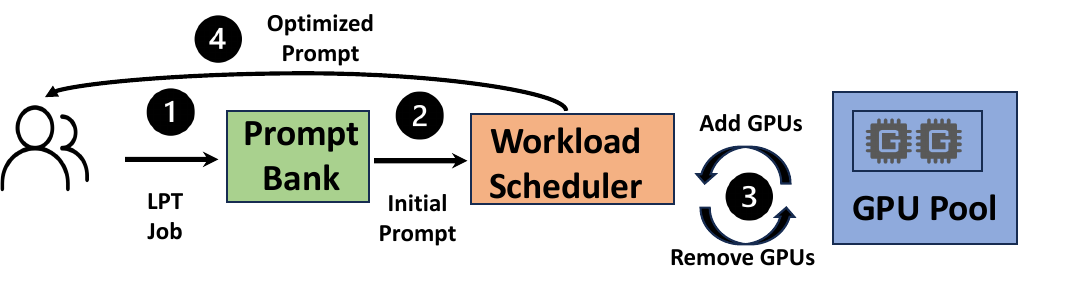}
\vspace{-10pt}
\caption{\small The workflow of \SysName{}. It consists of two key components: (1) The Prompt Bank identifies an effective initial prompt for an incoming LPT job at a minimal cost; (2) The Workload Scheduler dynamically adds GPUs from the GPU pool for each LPT job to reduce SLO violation while minimizing resource costs.
}
\label{fig:scheduler-architecture}
\vspace{-10pt}
\end{figure}

\begin{table}[t]
    \scriptsize
    \caption{\small{Job attributes description in \SysName{}.}}
    \vspace{-10pt}
    \label{tab:api}
    \centering
    \begin{tabular}{|m{1.8cm}|m{5cm}|}
        \hline
        \textbf{Attributes} & \textbf{Description} \\
        \hline
        \texttt{Model} & The LLM model name. \\
        \hline
        \texttt{Termination Condition} & The job completion criteria, including a maximum number of iterations and an accuracy target. \\
        \hline
        \texttt{Deadline} & The anticipated time by which the LPT workload should be completed. \\
        \hline
        \texttt{Dataset} & A path (e.g., AWS S3) where data samples are stored. \\
        \hline
        \texttt{Hyperparam} & Including initial prompt and parameters such as batch size, optimization algorithm.  \\
        \hline
        \hline
        \texttt{Prompt} & The optimized prompt. \\
        \hline
    \end{tabular}
\end{table}

\subsection{System Overview}
\label{subsec:system-system-overview}
\SysName{} contains two key components: the Prompt Bank leverages \emph{prompt reusing} to identify effective initial prompts for incoming LPT jobs (\S\ref{sec:prompt_bank}); the Workload Scheduler harnesses \emph{runtime reusing} to rapidly allocate GPUs to each LPT job, maintaining the SLO requirement while reducing resource costs (\S\ref{sec:prompt_sched}). 

Figure~\ref{fig:scheduler-architecture} shows the workflow of \SysName{}. First, the user submits an LPT job to the service provider ({\tiny\encircle{\normalsize1}}). The Prompt Bank identifies the effective initial prompt for this job ({\tiny\encircle{\normalsize2}}). Next, the Workload Scheduler dynamically adds/removes GPUs from/to the GPU pool based on the GPU demand of incoming traffic. The Workload Scheduler also dynamically adjusts the amount of GPUs for each job periodically ({\tiny\encircle{\normalsize3}}). Finally, the LPT service provider returns the optimized prompt to the user upon the LPT job completion ({\tiny\encircle{\normalsize4}}).

A job in \SysName{} is equivalent to an RPC request sent by an LPT service user followed by the RPC response from the system. 
Table~\ref{tab:api} summarizes the job attributes and descriptions. The first five attributes are job parameters specified by users. The last parameter is the response with an optimized prompt that the system returns to the user. The SLO of a job is defined as the maximum time during which the LPT job meets the termination condition.

\subsection{Prompt Bank}
\label{sec:prompt_bank}
The Prompt Bank realizes \emph{prompt reusing} to improve the ITA performance of incoming LPT jobs. It contains a set of prompts shared by all LPT jobs for selection as their initial prompts. We aim to \textit{balance} the speedup benefits of identifying initial prompts and latency cost of the query. To this end, we engineer the Prompt Bank as a query engine with a two-layer data structure. It enables efficient lookup operations for new LPT jobs and facilitates the seamless insertion and replacement of new initial prompt candidates. Below we detail the process of constructing the data structure and performing lookup, insertion, and replacement operations on it. The notations used in this section are defined in Table~\ref{tab:summary-notation-prompt-bank}. 


\begin{table}[h]
    \vspace{-5pt}
	\caption{Summary of Notations in the Prompt Bank.}
        \vspace{-10pt}
	\label{tab:summary-notation-prompt-bank}
	\centering
    \begin{adjustbox}{width=0.48\textwidth}
	\begin{tabular}{cl}
		\toprule
		\textbf{Sym.} & \textbf{Definition} \\
		\midrule
            $\mathcal{D}_{\text{eval}}$ & The evaluation dataset \\
            $d^{\text{in}}_i$ & The input query sample \\
            $d^{\text{tgt}}_i$ & The target response sample \\
            $\text{concat}$ & The operation to concatenate two text sequences \\
            $\mathcal{L}$ & The loss between the output and target sample \\
            $C$ &  The total number of prompt candidates in the Prompt Bank \\
            $K$ & The number of clusters for algorithm K-medoid \\
            $C_{\text{sim}}$ & The cluster with the representative prompt that is closest to the new initial prompt \\
            \bottomrule
	\end{tabular}
    \end{adjustbox}
        \vspace{-15pt}
\end{table}

\begin{figure}[t]
  \centering
  \includegraphics[width=.48\textwidth]{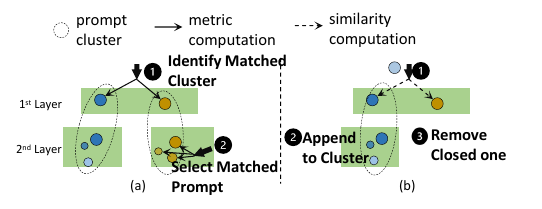}
  \vspace{-25pt}
\caption{\small The illustration of performing (a) lookup, and (b) insertion \& replacement on the two-layer data structure.}
\vspace{-15pt}
\label{fig:two-stage-evaluator}
\end{figure}

\subsubsection{Data Structure Construction}
We first assemble thousands of prompt candidates from public sources~\cite{prompthero,awesome_chatgpt_prompts} into a comprehensive set, which can maximize the likelihood of selecting effective initial prompts. To identify an effective initial prompt for a given LPT job, a brute-force search over the entire prompt candidate set is computationally intensive, often taking hours. Our empirical study (Figure~\ref{fig:promt-sim} in \S\ref{sec:evaluation}) and existing LLM research~\cite{su-etal-2022-transferability} demonstrate the prevalence of prompt similarity. This provides an opportunity to exclude unnecessary assessment of extensive poor prompt candidates and improve query efficiency~\cite{awesome_chatgpt_prompts}.

To this end, we build a two-layer data structure for the prompt candidate set. Inspired by~\cite{ModelKeeper}, we divide all the prompt candidates into clusters based on their activation feature similarity. We begin by using an LLM (e.g., Vicuna-7B) to extract the activation features of each prompt candidate. Then, we measure the prompt similarity based on the cosine distance between activation features. We also discuss other similarity metrics in~\S\ref{sec:implementation-details-prompt-bank}. Finally, we adopt K-medoid clustering to group prompts with similar activation features into one cluster. Figure~\ref{fig:two-stage-evaluator} (a) illustrates an example of this data structure. The first layer retains each cluster's medoid, further referred to as the \textit{representative prompt} of the cluster. The second layer stores each prompt within these clusters. Hereafter, we detail how to perform the lookup, insertion \& replacement operations. 


\subsubsection{Lookup}
The lookup operation aims to identify an effective initial prompt for a given LPT job on this two-layer data structure. For each initial prompt candidate \( p \), we introduce a metric \( \mathtt{score}(p) \), which is computed as the average loss on evaluation samples without requiring additional tuning on the training samples. We formulate \( \mathtt{score}(p) \) as follows: 
\begin{equation}
    \label{eq:score}
    \mathtt{score}(p) = \frac{1}{\mathcal{D}_{\text{eval}}}\sum_{(d^{\text{in}}_i, d^{\text{tgt}}_i) \in \mathcal{D}_{\text{eval}}} \mathcal{L}\left(\text{concat}(p, d^{\text{in}}_i), d^{\text{tgt}}_i\right). 
\end{equation}
A smaller \( \mathtt{score} \) value indicates a better initial prompt. We only use a small number of evaluation samples (e.g., 16) for prompt assessment. This requires minimal effort for labeling if the evaluation dataset is missing. Without performing any tuning, we can select the prompt with the minimal score as the most effective one. 





The two-layer data structure facilitates the reduction of the number of prompt candidates needed to perform the metric computation in Eqn.~\ref{eq:score}. Figure~\ref{fig:two-stage-evaluator} (a) illustrates the process of lookup operation. First, we identify the matched cluster by computing each representative prompt's \(\mathtt{score} \) at the first layer. We identify the cluster with the lowest \( \mathtt{score}\). Next, we select the matched initial prompt by calculating the \( \mathtt{score}\) for each prompt of the matched cluster at the second layer. We pick up the prompt with the lowest \( \mathtt{score} \) as the optimal one. Assuming that each cluster contains the same number of prompt candidates, this two-layer data structure reduces the number of metric computations from $C$ to $K + C/K$. Ideally, the minimal number of metric computations is $2\sqrt{C}$ when the optimal cluster is $K=\sqrt{C}$. Empirically, the two-layer data structure can reduce the overhead of the lookup operations by up to 40\( \times \) while retaining the performance (\S\ref{sec:eval-ablation-studies}). 


\subsubsection{Insertion \& Replacement}
When the service provider inserts a new initial prompt candidate, Figure~\ref{fig:two-stage-evaluator} (b) shows the process of the insertion and replacement operation. First, we identify a similar cluster. We extract the activation features of the new candidate and measure the cosine distance of activation features between the new candidate and each cluster's representative candidate at the first layer. Different from the lookup operations, we do not involve metric computations (Eqn.~\ref{eq:score}) in this step. The cluster that attains the minimal cosine distance is denoted as \( C_{\text{sim}}\). Second, we append this initial prompt into \( C_{\text{sim}}\) at the second layer. Third, the replacement operation is triggered when the number of initial prompt candidates exceeds the threshold (e.g., 3000) after insertion. We need to select one prompt candidate to remove it. To maximize the diversity of prompt candidates within the cluster, we choose the prompt candidate that has the minimal cosine distance to the representative prompt of \( C_{\text{sim}}\) and remove it to realize the replacement.


\subsubsection{Two-layer Structure Discussion}
\label{subsec:two-layer-structure-discussion}
The prevalent similarities among prompts suggest that clustering similar prompts can avoid unnecessary score assessment with minor speedup benefit loss. The study in \S\ref{sec:eval-ablation-studies} indicates that a two-layer data structure can identify effective initial prompts within 10 seconds. Additionally, we construct a three-layer structure using K-medoid clustering, but encounter convergence issues with Vicuna-7B and experience exorbitant construction overhead (up to tens of minutes). A two-layer structure can be constructed in five minutes without convergence issues across different LLMs, making it a more suitable choice.

\subsection{Workload Scheduler} 
\label{sec:prompt_sched}
\begin{figure}
\centering
  \centering
  \includegraphics[page=1, width=0.5\textwidth]{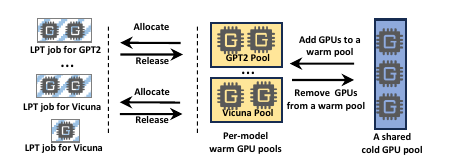}
\vspace{-15pt}
\caption{\small The Workload Scheduler consists of a single shared cold GPU pool and a set of per-LLM warm GPU pools. It rapidly allocates GPUs from the warm GPU pools to LPT jobs to optimize the SLO attainment. It also dynamically adjusts the number of GPUs added from the shared cold GPU pool to the warm GPU pools based on traffic and GPU availability.}
\vspace{-15pt}
\label{fig:workload-scheduler}
\end{figure}

The Workload Scheduler realizes \emph{runtime reusing} to mitigate the exorbitant GPU allocation overhead (\S\ref{sec:characterization-prompt-tuning}), thus reducing the SLO violation and minimizing the resource cost. Figure~\ref{fig:workload-scheduler} shows the overview of the Workload Scheduler, which manages two types of GPU pools, namely a single \emph{shared cold} GPU pool and a set of \emph{per-LLM warm} GPU pools. Each warm pool contains GPUs initialized to serve jobs for one specific LLM, i.e., each GPU has a pre-loaded PyTorch/CUDA runtime and LLM weights. The shared cold GPU pool contains GPUs without any pre-loaded GPU context\footnote{Although cloud providers are free to use the GPUs from the cold pool for any jobs operating in their datacenter, for simplicity, we assume that the size of the cold GPU pool is fixed and GPUs can be allocated without any delays in time.}.

Managing the per-LLM warm pools independently from the shared cold pool significantly reduces GPU allocation overhead without statically provisioning a large fixed-size cluster, as in ElasticFlow (\S\ref{sub:dl-training-limitations}). When the scheduler allocates GPUs to an LPT job from the corresponding warm pool, the job can start execution immediately, avoiding the delays of pre-loading the required runtime and LLM weights. Thus, the Workload Scheduler facilitates \emph{runtime reusing} of LPT jobs of the same LLM. Since many users use the same LLMs~\cite{llama-7b,vicuna,GPT-2}, the system can operate efficiently while minimizing the operational cost by keeping only a small number of warm pools with a minimal set of GPUs.  
Unlike the GPUs in the warm pools, the GPUs in the cold pool do not impose any cost, so the providers can allocate them to any service running in datacenters.

For each incoming job in the pending queue, the Workload Scheduler determines the number of GPUs to be allocated based on its SLO and predicted execution time. It then allocates GPUs from the warm pool corresponding to the LLM type defined in the job attributes. To secure the SLO compliance, we predict the upper bound of job execution time as a product of the number of maximum remaining iterations and maximum time cost per iteration under given allocated GPUs with additional GPU allocation overhead. LPT jobs release their allocated GPUs to the corresponding warm pools upon completion. The scheduler monitors each pool's GPU usage, adding more GPUs from the cold pool to the warm pools of the LLMs that experience high demand and removing excessive GPUs from the warm pools of the LLMs. 


Next, we detail two key resource allocation algorithms for LPT execution and one algorithm for the execution of Prompt Bank. Table \ref{tab:summary-notation-workload-scheduler} defines the notations used in this section.




\begin{table}[t]
	\caption{\small Summary of notations in the Workload Scheduler.}
        \vspace{-10pt}
	\label{tab:summary-notation-workload-scheduler}
	\centering
    \begin{adjustbox}{width=0.48\textwidth}
	\begin{tabular}{cl}
		\toprule
		\textbf{Sym.} & \textbf{Definition} \\
		\midrule
            $i$ & The index of LPT job \\
            $l$ & The index of LLM \\
            $k$ & The index of GPU in a warm GPU pool \\
            $a$ & The number of allocated GPUs \\
		$L$ & The number of LLMs. \\
		$R_l$ & The number of GPUs for each LLM $l$'s warm GPU pool \\
            $A$ & The number of allocated GPUs in a warm GPU pool for each job \\
            $B$ & The number of GPUs added from the cold GPU pool for each warm GPU pool \\
            $T^{\text{slo}}_i $ & The SLO of job $i$ \\
            $T_i^{\text{warm}}(a) $ & The estimated completion time of job $i$ when assigned with $a$ GPUs in a warm GPU pool \\
            $T^{\text{cold}}$ & The overhead of adding GPUs from the cold GPU pool to a warm GPU pool \\
            $\mathcal{P}$ & All pending LPT jobs \\
            $E_l$ & The list to store earliest timestamps for each GPU in the LLM $l$'s warm GPU pool \\
            $E$ & A list to store each LLM's $E_l$ \\
		\bottomrule
	\end{tabular}
    \end{adjustbox}
        \vspace{-15pt}
\end{table}

\subsubsection{GPU Allocation from a Warm Pool}
\label{sec:warm-GPU-allocator}
This algorithm optimizes the SLO attainment by determining the number of GPUs in the warm GPU pools allocated to each job in the pending queue. Upon an LPT job's arrival, the scheduler adds it to the pending queue. Then, the scheduler periodically adjusts the GPU allocation for each job in the queue, allocating more GPUs from the corresponding warm pool whenever needed to achieve the job's SLO. Algorithm~\ref{alg:warm-resource-allocator} illustrates this process. It starts by sorting LPT jobs in the pending queue based on their SLOs, and then progressively increases the number of allocated GPUs for each LPT job to meet its SLO until the warm pool is depleted (Lines~\ref{line:warm-start}-\ref{line:warm-end}).

\begin{algorithm}[t]
\footnotesize
\caption{GPU allocation from a warm pool.} \label{alg:warm-resource-allocator}
\begin{algorithmic}[1]
\State \textbf{Input:} $R_l$ that is the number of GPUs in the LLM $l$'s warm pool, $\mathcal{P}_l$ that is the pending queue for LLM $l$.
    \State \textbf{Output:} $A$ that is the number of GPUs allocated to each job in the pending queue. 
\\
\State Sort jobs based on $T^{\text{slo}}_i$ in the ascending order
\For{each job $i$ in $\mathcal{P}_l$
}
    \State Set initial allocation $A_i = 1$
    \While{$T_{i}^{\text{warm}}(A_i) > T_{i}^{\text{slo}}$ \textbf{and} $A_i \leq R_l $} \label{line:warm-start}
        \State $A_i = A_i + 1$ \textit{// \textcolor{blue}{Allocate $A_i$ GPU to the job}} 
    \EndWhile \label{line:warm-end}
    \If{$T_{i}^{\text{warm}}(A_i) \leq T_{i}^{\text{slo}}$} 
        \State $R_l = R_l - A_i$ \textit{// \textcolor{blue}{Update the number of GPUs in the warm GPU pool}}
    \Else 
        \State $A_i = 0$
    \EndIf
\EndFor
\end{algorithmic}
\end{algorithm}

\begin{algorithm}[t]
\footnotesize
\caption{GPU allocation from the cold pool.} \label{alg:cold-resource-allocator}
\begin{algorithmic}[1]
\State \textbf{Input:} $L$ LLM, pending queue with jobs $\mathcal{P}$, earliest timestamps of GPUs in the warm GPU pools $E$.
    \State \textbf{Output:} The number of allocated GPUs $B$ to each LLM's warm GPU pool.

\State
\State Sort $\mathcal{P}$ based on $T^{\text{slo}}_i$ in the ascending order 

\For{each job $i$ and the corresponding LLM $l$ in $\mathcal{P}$}
    \State \textit{// \textcolor{blue}{Assess if the system can meet the job's SLO by delay its execution}}
    \If {\textsc{DelaySchedulable}($E$, $i$, $l$)} 
        \State \textbf{continue} 
    \EndIf
    \State Set the initially allocated GPU number $A_i = 1$
    \State \textit{// \textcolor{blue}{Determine how many GPUs are needed to satisfy the job's SLO}}
    \While{$T_{i}(A_i) + T^{\text{cold}}_{l} >  T_{i}^{\text{slo}}$ \textbf{and} $T_{i}^{\text{slo}} < T^{\text{cold}} $ } \label{line:cold-start}
        \State $A_i = A_i + 1$ 
    \EndWhile \label{line:cold-end}

    \If{$T_{i}(A_i) + T^{\text{cold}}_{l} \leq  T_{i}^{\text{slo}}$}
    \State \textit{// \textcolor{blue}{Update the number of GPUs in each LLM's warm pool}} \label{line:others-start}
        \State $B_{l} = B_{l} + A_i$ 
        \State \textit{// \textcolor{blue}{Update the earliest timestamps of GPUs in the warm GPU pools.}} 
         \State Repeat $A_i$ times to push back $T_{i}^{\text{warm}}(A_i)+ T^{\text{cold}}_{l}$ into $E_l$. 
    \EndIf \label{line:others-end}
\EndFor

\State

\Function{DelaySchedulable}{$E$, $i$, $l$} \label{alg:func-start}
    \State $k$ = 1, $T^{\text{cur}}$ = current timestamp 
    \State Sort $E_{l}$ in the ascending order
    \While{$k \leq E_l.len $ \textbf{and} $T_{i}(k)- T^{\text{cur}} + E_{l,k} >  T_{i}^{\text{slo}}$   } \label{line:delay-start}
        \State $k = k + 1$
    \EndWhile\label{line:gpu-end}
    \If {$k < E_l.len$ \textbf{and} $T_{i}(k) - T^{\text{cur}} + E_{l,k}  \leq  T_{i}^{\text{slo}}$} 
        \State $E_{l,1:k} = T_{i}(k) + E_{l,k}-T^{\text{cur}}$
        \State Sort $E$ in the ascending order
        \State \textbf{return} True
    \EndIf
    \State \textbf{return} False
\EndFunction \label{alg:func-end}

\end{algorithmic}
\end{algorithm}

\subsubsection{GPU Allocation from the Cold Pool}

\label{sec:cold-gpu-allocator}

The Workload Scheduler can periodically add and remove GPUs from the cold GPU pool to the corresponding warm GPU pool, following the demand for the corresponding LLM. The main objective of the algorithm is to ensure each warm pool has the minimum number of GPUs required to ensure that the jobs can achieve their SLOs while minimizing the resource cost, which is proportional to the number of GPUs used by the jobs and present in the warm pools. Hence, the algorithm prioritizes jobs with shorter SLOs, delaying the execution of the jobs with longer SLOs and the jobs projected to miss SLOs.

Algorithm~\ref{alg:cold-resource-allocator} details the steps that allocate GPUs from the cold pool to the warm pools. First, the algorithm sorts all pending jobs based on its SLO. Second, it identifies the LPT job $i$, scheduling of which can be delayed while still meeting its SLO by calling the \textsc{DelaySchedulable} function (Line~\ref{alg:func-start}-\ref{alg:func-end}). Third, if the system cannot meet the job's SLO, the algorithm progressively allocates more GPUs from the cold GPU pool to the job until it can ensure that SLO is met. The algorithm takes the GPU allocation overhead $T_{l}^{\text{cold}}$ into account while determining whether the system can meet the job's SLO (Line~\ref{line:cold-start}). Last, if the system can meet the job $i$'s SLO, the algorithm accumulates the number of added GPUs from the cold GPU pool to the corresponding warm GPU pool (Line~\ref{line:others-start}-\ref{line:others-end}).

The \textsc{DelaySchedulable} function determines if a job's SLO can be met by delaying its execution to a future moment when enough GPUs would be released by completing jobs to the warm pool instead of immediately adding more GPUs to the warm pool. We use $E_{l,k}$ to record the earliest timestamp when $k$ GPUs in a warm GPU pool for LLM $l$ are available. This information is obtained from predicting the completion time of each running LPT job, along with the subsequent release of GPUs to the respective warm pool. Additionally, to reduce the resource cost, the Workload Scheduler removes the GPUs from a warm pool if they do not serve any jobs for a time window, the size of which is set to one minute (\S\ref{sec:eval-ablation-studies}). 



\noindent\textbf{Why \textsc{DelaySchedulable} function.} Many SLO-aware resource allocation policies~\cite{elasticflow,INFaaS,INFless} expect to schedule jobs promptly to ensure SLO compliance for jobs. Delaying the execution of the jobs might risk the SLO violation. Owing to the speedup benefits provided by the Prompt Bank, many LPT jobs are completed earlier, leaving many GPUs idle. The Workload Scheduler takes advantage of this by strategically delaying the execution of LPT requests without requiring launching additional GPU resources. This allows the system to meet SLOs for more LPT requests with fewer GPUs.

\subsubsection{Latency Budget for Prompt Bank}
The Workload Scheduler needs to allocate GPUs to perform the Prompt Bank. Despite we support the sequential execution of Prompt Bank and LPT (\S~\ref{sec:implementation-details-prompt-bank}) and reduce the overhead of the Prompt Bank within 10 seconds, it is possible that this overhead compromises SLO compliance for short requests. We empirically observe that the Prompt Bank can yield a 1.2-4.7\(\times\) speedup compared to the induction initialization~\cite{ye2023prompt}, an automatic prompt initialization baseline (detailed in \S\ref{sec:experimental-setup}). Therefore, we set a latency budget of 20\% of the latency SLO to execute the Prompt Bank, ensuring that the minimum speedup benefits still outweigh the overhead of the Prompt Bank.

\section{Implementation}
\label{sec:implementation}






\subsection{Multi-GPU Execution}
\label{sec:preemptive-elastic-execution}
We implement LPT jobs with ~2000 lines of Python code atop Transformers 2.4.1 and PyTorch 2.1 and deploy them as containerized GPU Knative functions to pre-load the LPT runtime and LLM weights in the GPU. Each Knative function accepts a set of parameters described in Table~\ref{tab:api} and responds to users with the optimized prompt. We adopt the prompt-tuning algorithm in~\cite{TEXTBOX}. Note that \SysName{} is general and can support other implementations of LPT jobs and algorithms.

An LPT job demands multiple function instances to deliver multi-GPU execution. We implement the multi-GPU execution atop LambdaML~\cite{lambdaml}, which employs Memcached as the storage channel to realize the synchronous cross-GPU communication between function instances. Each function instance belonging to an LPT job is assigned an IP address and port to connect with other function instances, incurring at most a 2-second overhead. The storage channel incurs negligible communication overhead due to its small size. 

\subsection{Prompt Bank}
\label{sec:implementation-details-prompt-bank}
We implement the Prompt Bank with $\sim$1000 lines of Python code atop Transformers 2.4.1 and PyTorch 2.1. It is also deployed as a Knative function with one GPU, which accepts parameters, including the dataset and initial prompt described in Table~\ref{tab:api}, and returns the optimized initial prompt for subsequent prompt-tuning.

\noindent\textbf{Offline Phase.} For each LPT job, we use the corresponding LLM to extract the activation features of gathered prompts and empirically set the number of clusters in the two-layer data structure as 50. Moreover, we employ Scipy 1.10.1 to execute K-medoid clustering. Despite exploring alternative distance metrics, including Manhattan and Euclidean distances, we encounter convergence issues. The lack of convergence may stem from imbalances in the numerical value scales within the activation features of various prompts. The storage size remains under 5 GB for each LLM. We have detailed the insertion and replacement operation in \S\ref{sec:prompt_bank}. If the service provider introduces a new LLM, it needs to re-extract the activation features of all gathered prompts to construct the two-layer data structure.



\noindent\textbf{Online Phase.} The Workload Scheduler utilizes the latency budget to decide whether to perform the Prompt Bank for incoming request. We notice that the implementation and runtime of the Prompt Bank and LPT can be shared. Hence, we incorporate the Prompt Bank into the corresponding LPT job. In other words, the Prompt Bank and LPT job run sequentially on their assigned GPUs. 



\subsection{Workload Scheduler} 
\noindent\textbf{Pre-loaded Runtime.} Each job requires multiple function instances for multi-GPU execution. Knative provides an autoscaling mechanism to maintain function instances serving future requests. 


\noindent\textbf{GPU Allocation from a Warm Pool.} This algorithm aims to perform rapid GPU allocation from a warm pool to an LPT job. Hence, we conduct the round-based GPU allocation every 50 milliseconds, which is negligible compared to minutes-level latency SLO. It operates within the distributed control plane of Knative to assign GPUs in the warm GPU pools to corresponding LPT jobs. 



\noindent\textbf{GPU Allocation from a Cold Pool.} This algorithm is implemented inside the distributed control plane of Knative, and the interval is set as 50 milliseconds to add and remove GPUs for warm pools promptly. Moreover, this algorithm tracks the profiled information, including the allocation overhead for each LLM and the job throughput. It then continuously updates this information to the scheduler to avoid high estimation errors. 

\section{Evaluation}
\label{sec:evaluation}

\begin{figure}
\begin{subfigure}{.5\textwidth}
    \centering
  \includegraphics[width=.85\textwidth]{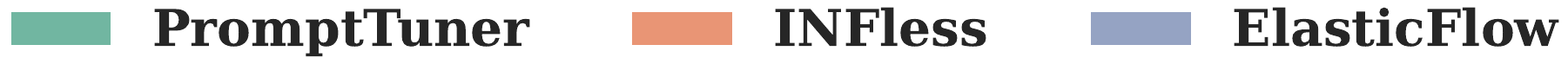}
\end{subfigure}
\begin{subfigure}{.24\textwidth}
  \centering\includegraphics[width=\textwidth]{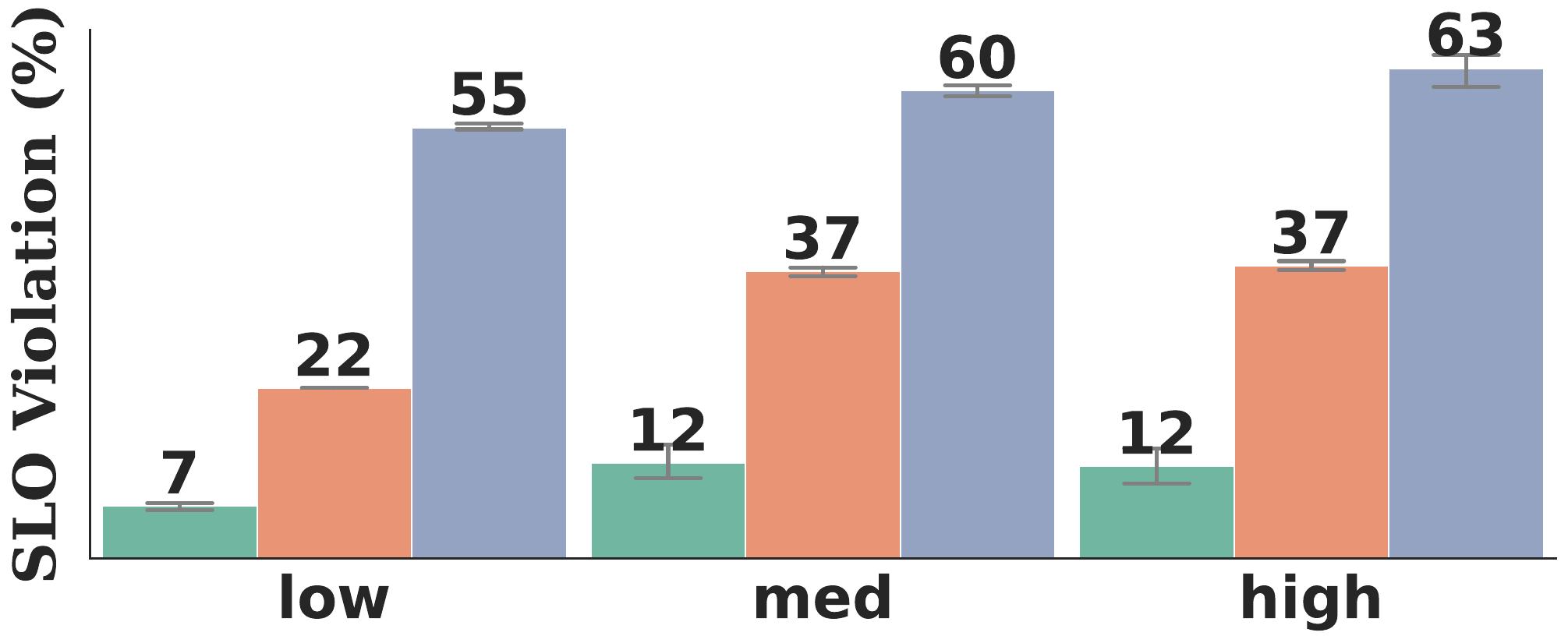}
  \caption{SLO Violation vs. Load.}
  \label{fig:e2e-density}
\end{subfigure}%
\begin{subfigure}{.24\textwidth}
\centering\includegraphics[width=\textwidth]{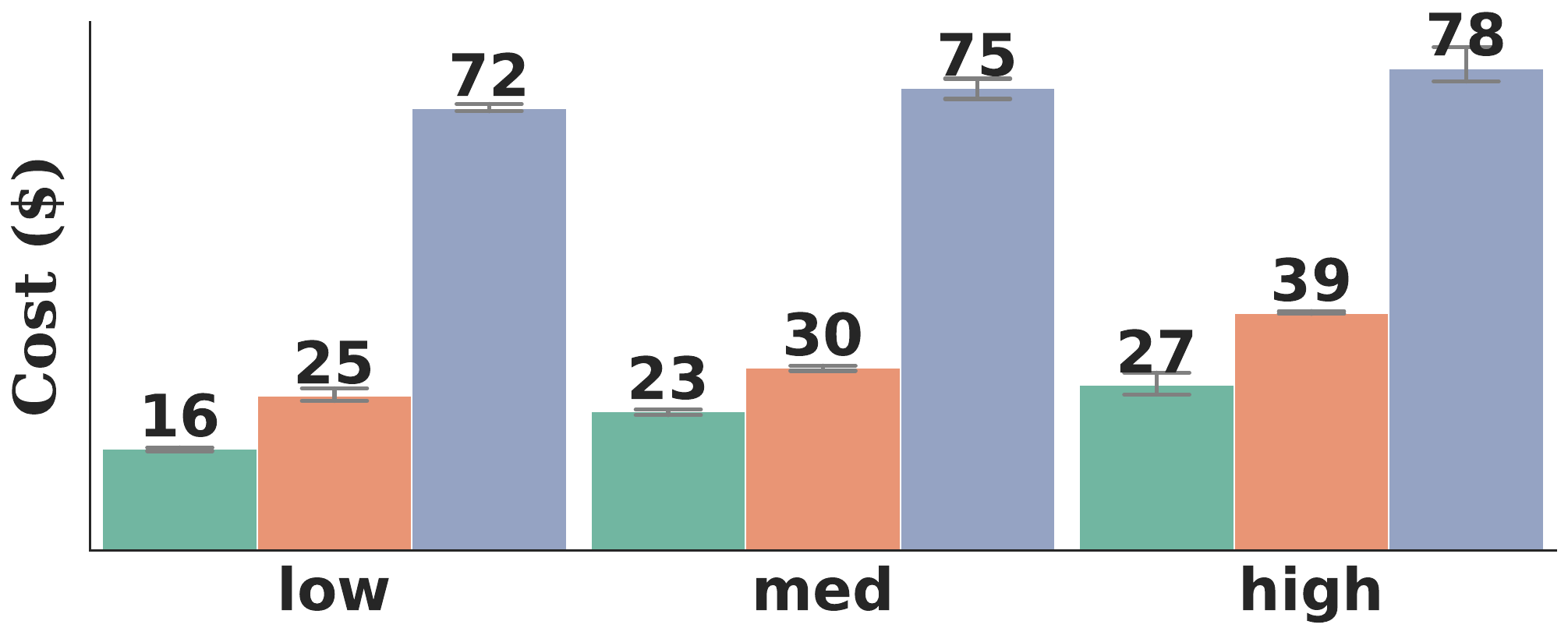}
  \caption{Cost vs. Load.}
  \label{fig:e2e-density-cost}
\end{subfigure}
\begin{subfigure}{.23\textwidth}
  \centering
  \includegraphics[width=\textwidth]{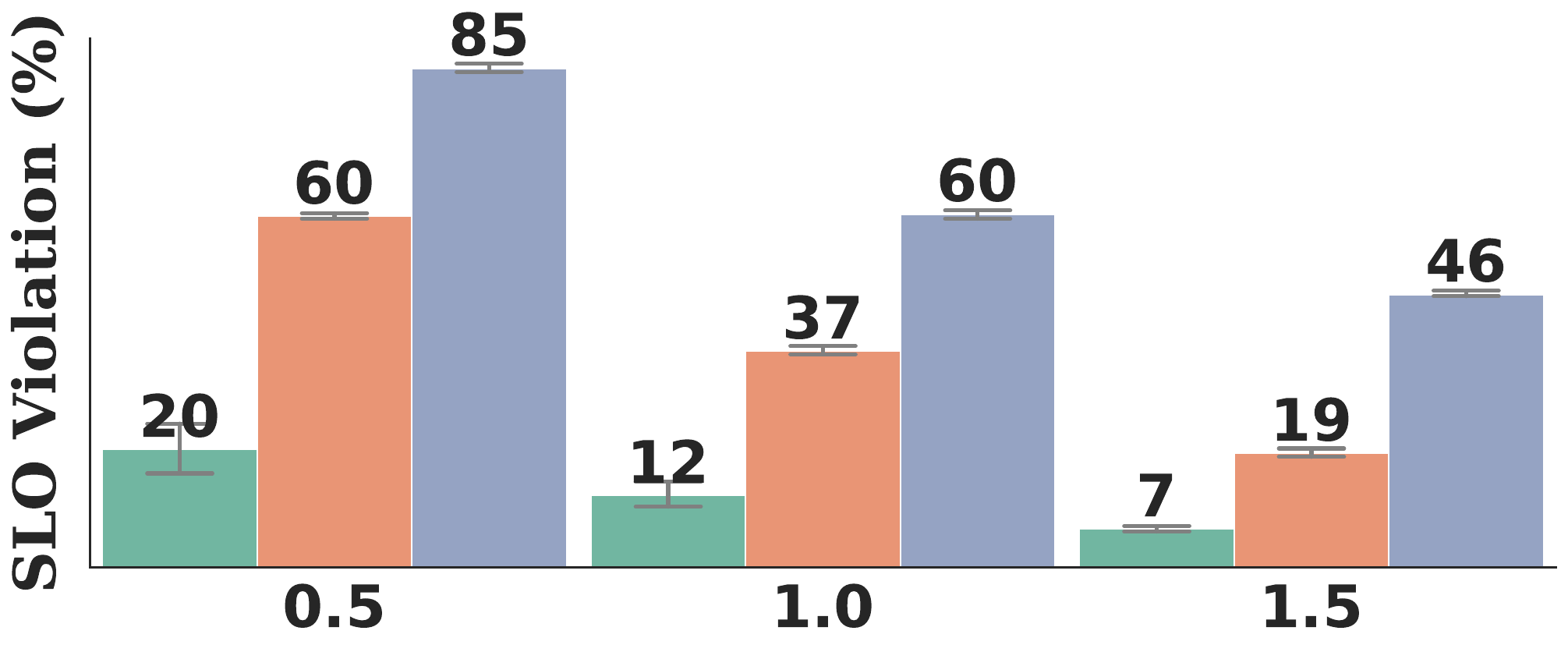}
  \caption{SLO Violation vs. Emergence.}
  \label{fig:e2e-slo}
\end{subfigure}
\begin{subfigure}{.23\textwidth}
  \centering
  \includegraphics[width=\textwidth]{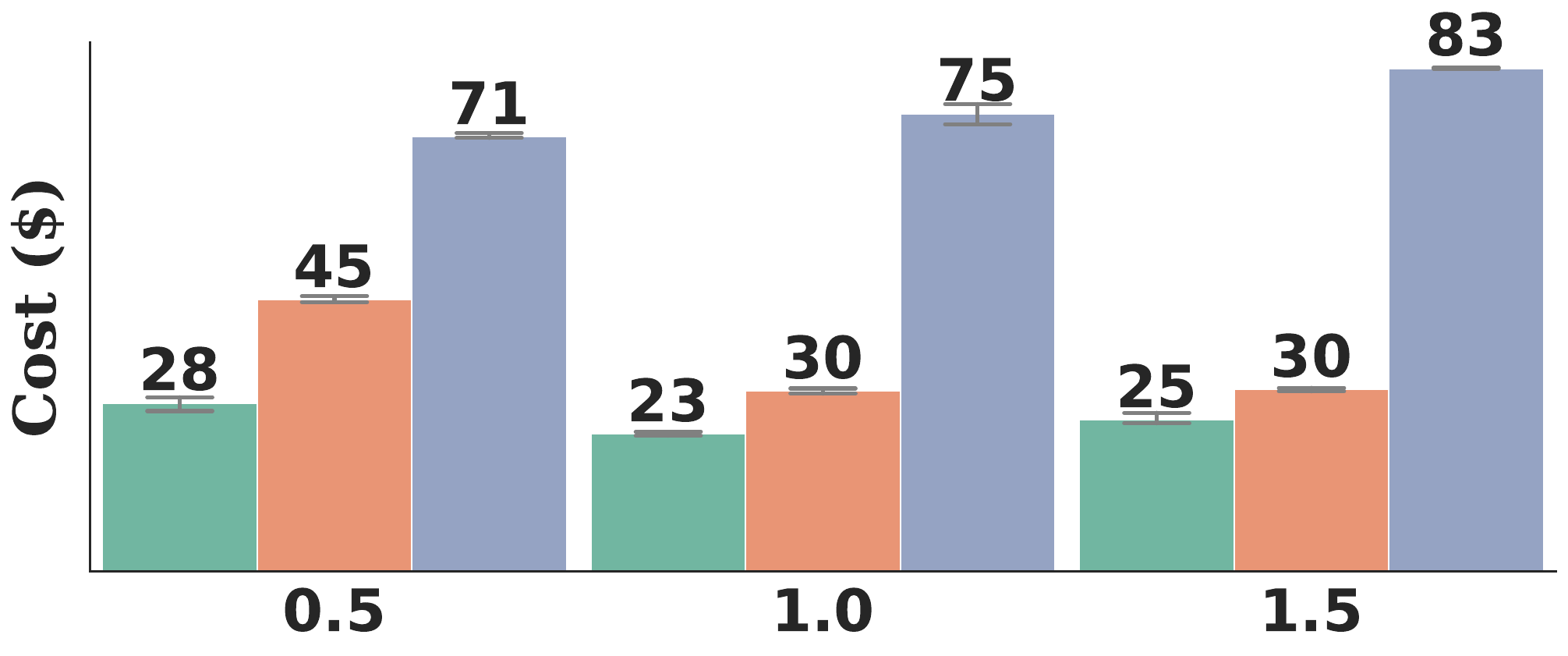}
  \caption{Cost vs. SLO Emergence.}
  \label{fig:e2e-slo-cost}
\end{subfigure}
\vspace{-5pt}
\caption{\small End-to-end performance under different loads (a-b) and different SLO emergence (c-d).}
\vspace{-15pt}
\label{fig:e2e}
\end{figure}

\subsection{Experimental Setup}
\label{sec:experimental-setup}

\noindent\textbf{Testbed.}
We set up \SysName{} in a physical GPU cluster. Each GPU server has eight NVIDIA A100-80GB GPUs and one 200Gbs HDR InfiniBand. It features an Intel Xeon 8369B 2.90GHz CPU with 64 cores, 256 GB RAM, and PCIe-III. \SysName{} provisions at most 4 GPU servers. We adopt Memcached 1.5.22 to set up an Elastic Cache service for communication among GPU servers. 
\noindent\textbf{Workload Construction.} 
We evaluate three representative LLMs (GPT-Base, GPT-Large, Vicuna-7B) on 12 datasets, as shown in Table~\ref{tab:llm-tasks}. To further increase the diversity of LPT workloads, we sample each dataset into ten exclusive partitions and construct 120 tasks for each LLM. For each LPT task in Table~\ref{tab:llm-tasks}, we measure the average accuracy over 20 initial prompts randomly selected from the Prompt Bank as the target accuracy. This primarily ensures that the evaluated LPT jobs, using different initial prompts in the prompt sensitivity analysis of \S\ref{sec:characterization-prompt-tuning}, can reach such accuracy. We also evaluate LLaMA-30B and Qwen7B-R1\footnote{We use soft prompt tuning algorithm~\cite{li2021prefix} and task GSM8K~\cite{cobbe2021gsm8k} to evaluate Qwen7B-R1. The selected initial textual prompt is positioned before the soft prompt.} to present the system efficiency  in the context of large-scale models and long-sequence inputs.


In our experiments, we sample three 20-minute LPT traces from a data center to construct low (41/55/42), medium (77/71/65), and high (99/85/76) loads for different LLMs (GPT2-B/GPT2-L/V7B). We sample 59 and 70 requests for LLaMA-30B and Qwen7B-R1 as medium load. These traces include the submission time, the number of allocated GPUs, and the duration of each LPT job. The job durations vary from a few seconds to several minutes. We follow the minute granularity of the submission time attribute to invoke the request with an exponential distribution. The product of the job duration and number of allocated GPUs is used to assign LPT tasks for each job. It randomly chooses one task in Table~\ref{tab:llm-tasks} to match the GPU time of such a job. We set each job's SLO as its duration multiplied by a parameter \( S \) added by the resource allocation overhead. We denote \(S\) as SLO emergence. A small \(S\) indicates a more emergent SLO.

\begin{table}[t]\centering
\caption{LPT tasks and targeted accuracy: [B] and [R] refer to the \texttt{bleu} score and \texttt{rouge} score respectively.}\label{tab:llm-tasks}%
\vspace{-10pt}
\resizebox{0.9\linewidth}{!}{
\begin{tabular}{lrc|lrc}\toprule
\textbf{Task Description} &\textbf{Dataset} &\textbf{Accuracy} &\textbf{Task Description} &\textbf{Dataset} &\textbf{Accuracy} \\\midrule
\multirow{2}{*}{Dialog} & DA~\cite{TEXTBOX} & 54 [B] &\multirow{3}{*}{Summarization} &CNNDM ~\cite{TEXTBOX} & 34 [B] \\
&PC~\cite{PC} & 19 [B] & &SAMSUM~\cite{SAMSUM} & 46 [B] \\\cmidrule{1-3}
\multirow{2}{*}{Question Answer} &COQAQG~\cite{COQAQG} & 51 [B] &  &XSUM~\cite{TEXTBOX} &  40 [B] \\ \cmidrule{4-6}
&QUORA~\cite{QUORA} & 21 [B]  &\multirow{3}{*}{Story Generation} &CMV~\cite{CMV} & 26 [R] \\\cmidrule{1-3}
\multirow{2}{*}{Text Generation} &WIKIBIO~\cite{TEXTBOX} & 70 [R] & &WP~\cite{WP} & 20 [R] \\
&WIKIP~\cite{WIKIP} & 22 [R] & &ROC~\cite{ROC} & 25 [R] \\
\bottomrule
\end{tabular}}
\vspace{-10pt}
\end{table}

\noindent\textbf{Baselines.} \SysName{} is the first SLO-aware system for LPT workloads. We choose two state-of-the-art DL cluster management systems as the baselines: (1) \textbf{INFless}~\cite{INFless}: this is an efficient SLO-aware and cost-effective system for DL inference. It supports traffic-based autoscaling and runtime reusing. To ensure a fair comparison, we reinforce INFless with the multi-GPU execution and Prompt Bank. (2) \textbf{ElasticFlow}~\cite{elasticflow}: this is an SLO-aware DL training system. It dynamically adjusts the number of GPUs for each job. However, it does not support runtime reusing. The Prompt Bank is also incorporated into ElasticFlow. 

To evaluate the quality of initial prompts from the Prompt Bank, we consider two baselines: (1) \textbf{Ideal}: this is the prompt with the best ITA performance. For easy computation, we use \( \mathtt{score} \) to shortlist 20 prompts and select the best one based on their ITA performance. However, it is computationally infeasible in practice. (2) \textbf{Induction}~\cite{ye2023prompt}: it is an automatic prompt initialization method that leverages a set of demonstrative examples to guide the LLM to generate an appropriate initial prompt. However, it only works for simple tasks, and the LLM should possess strong capabilities.

\noindent\textbf{Evaluation Metrics.} We consider two evaluation metrics: (1) the ratio of workloads that meet the SLOs. We use the SLO violation as the metric. (2) The total resource cost. We estimate the cost based on the price of the AWS \texttt{p4de.24xlarge} instance. The storage costs are billed on GB/hour (AWS elastic cache). We take the minimal possible price for storing transferred data, accounting for the small communication time. For the Prompt Bank, we choose ITA to demonstrate the high quality of selected initial prompts.

\begin{figure}
\begin{subfigure}{.42\textwidth}
    \centering
  \includegraphics[width=.65\textwidth]{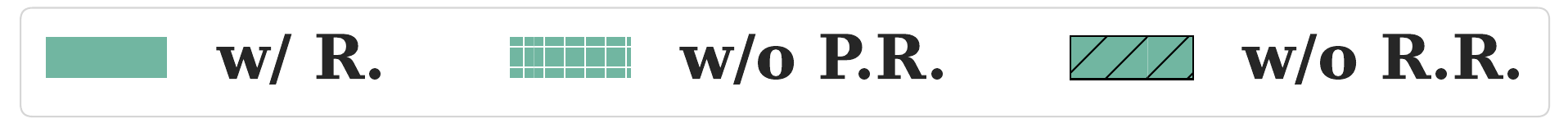}
\end{subfigure}
\begin{subfigure}{.23\textwidth}
  \centering    
  \includegraphics[width=\textwidth]{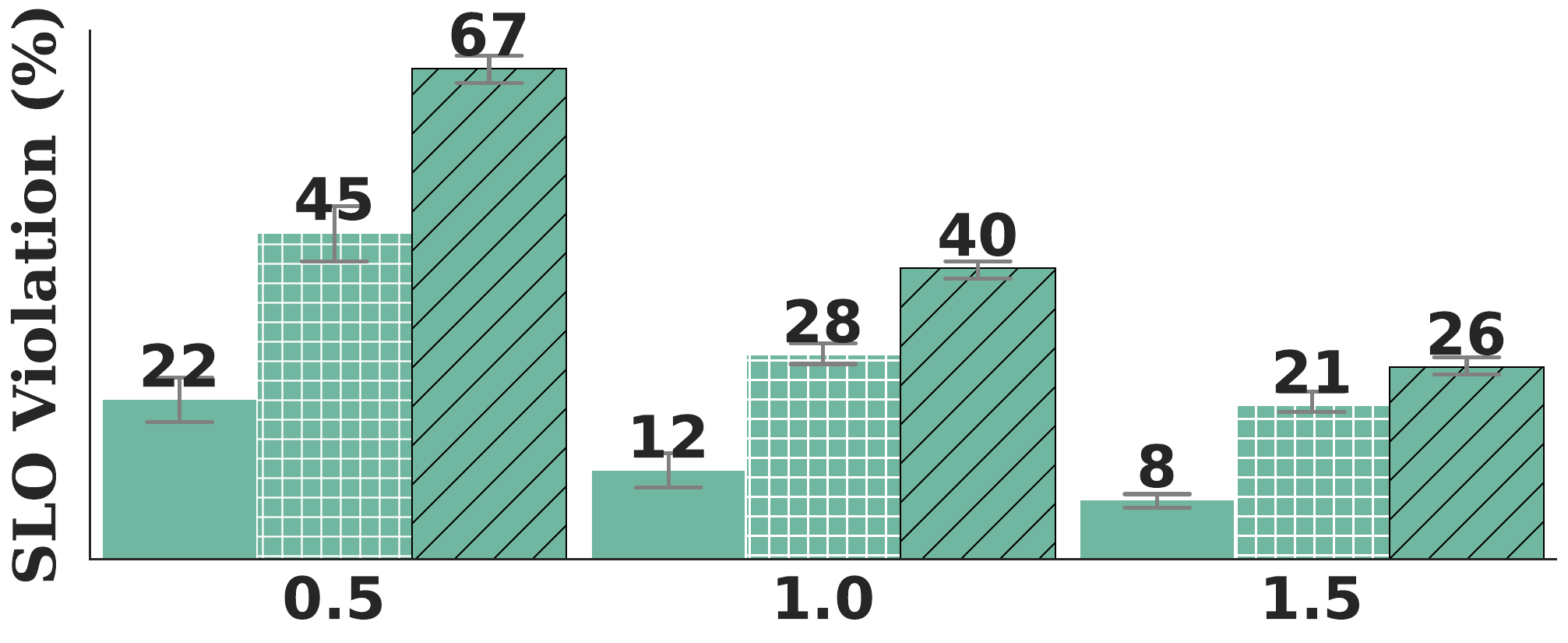}
  \caption{SLO Violation vs. Sharing.}
  \label{fig:prompt-sharing-slo}
\end{subfigure}
\vspace{2pt}
\begin{subfigure}{.23\textwidth}
  \centering
  \includegraphics[width=\textwidth]{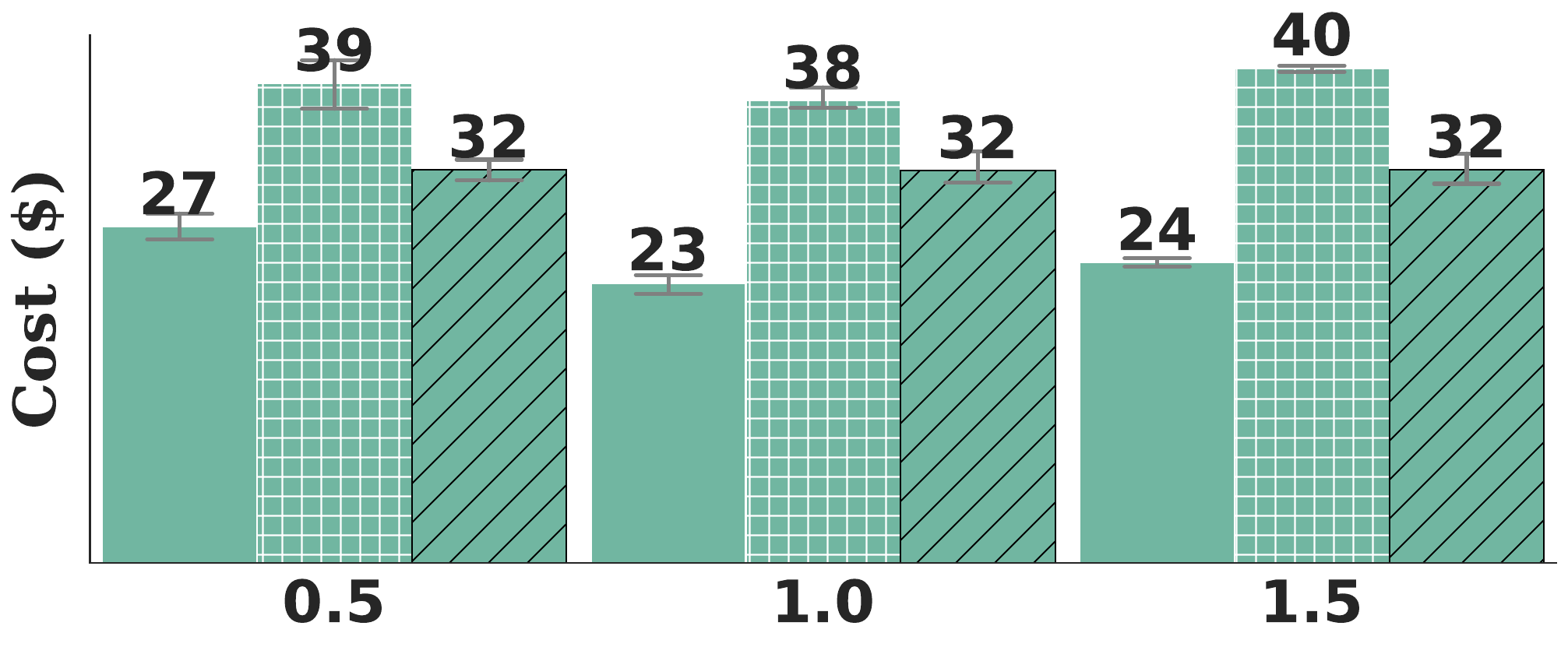}
  \caption{Cost vs. Sharing.}
  \label{fig:runtime-sharing-slo}
\end{subfigure}
\begin{subfigure}{.5\textwidth}
    \centering
  \includegraphics[width=.5\textwidth]{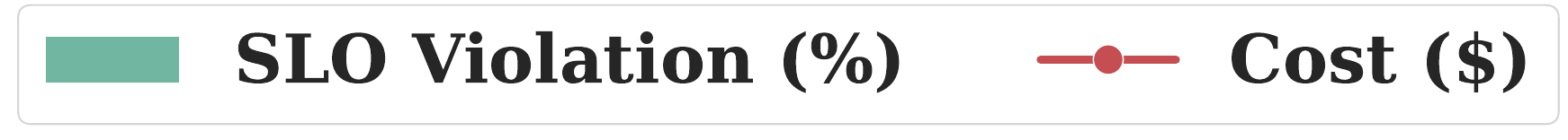}
\end{subfigure}
\begin{subfigure}{.23\textwidth}
  \centering
  \includegraphics[width=\textwidth]{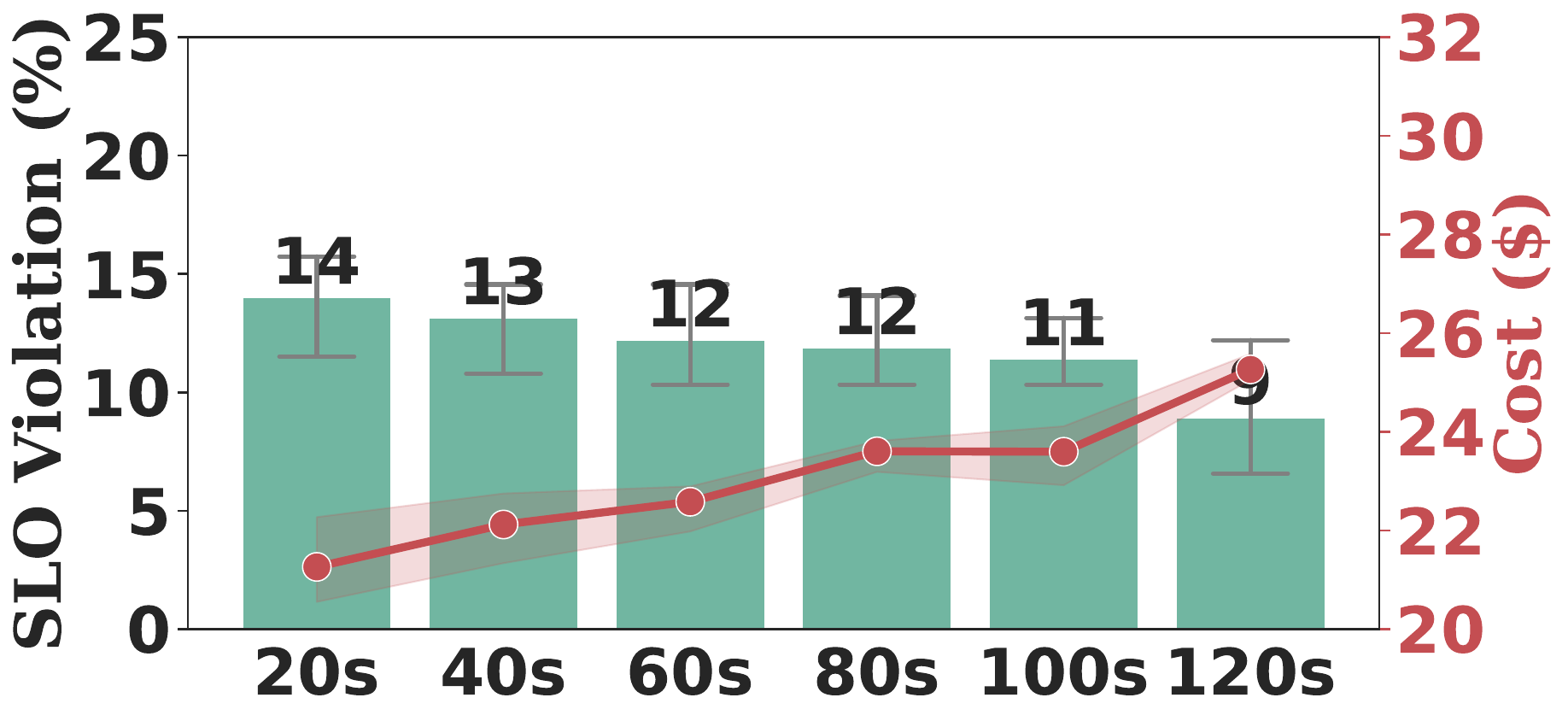}
  \caption{Window size.}
  \label{fig:window-time}
\end{subfigure}
\begin{subfigure}{.23\textwidth}
  \centering
  \includegraphics[width=\textwidth]{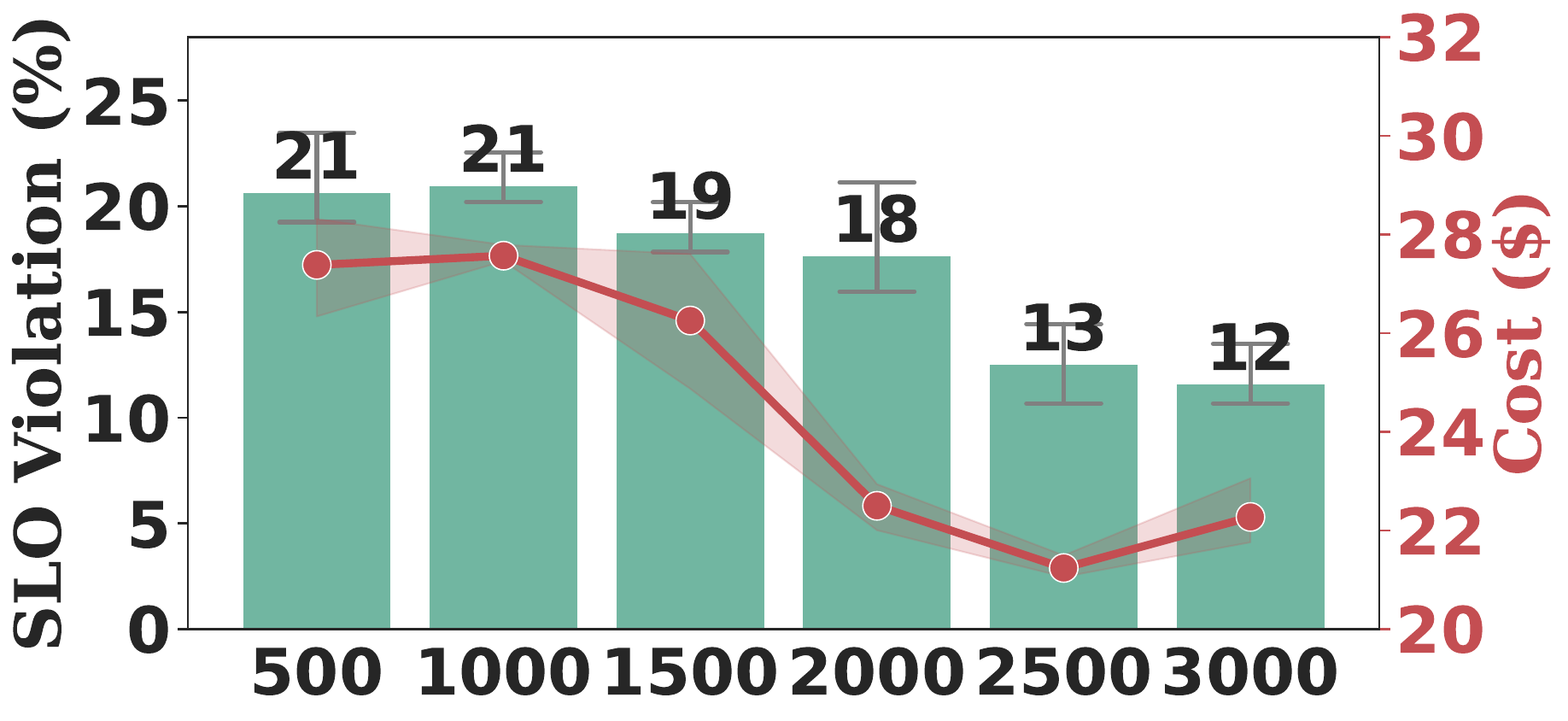}
  \caption{Bank size.}
  \label{fig:vary-prompt-bank-size}
\end{subfigure}
\vspace{-10pt}
\caption{\small Feature evalutions: (a-b) The impact of prompt reusing (P.R.) and runtime reusing (R.R.) on SLO violation and  cost over different SLO levels. (c-d) SLO violation and cost of \SysName{} under varying window sizes (c) and prompt bank sizes (d).}
\vspace{-15pt}
\label{fig:impact-of-systems}
\end{figure}

\subsection{End-to-end Performance}
\label{sec:system-in-cluster}
We compare the end-to-end performance of \SysName{} with two baselines (INFless and ElasticFlow) under various environments in a physical cluster. Our empirical evaluation in Figure \ref{fig:e2e} simultaneously serves requests for three LLMs in this experiment. First, Figures~\ref{fig:e2e-density} and~\ref{fig:e2e-density-cost} present the SLO violation and cost of these systems under different job loads, respectively. \SysName{} achieves 15-25\% SLO violation reduction compared to INFless and 48-51\% SLO violation reduction compared to ElasticFlow. Interestingly, the increased loads provide more opportunities to perform \textit{runtime reusing}. Thus, the SLO violation does not increase significantly from medium to high loads. The heavy job load increases the SLO violation and cost, and \SysName{} demonstrates higher superiority than baselines under heavier job loads.

Second, we explore the SLO violation and cost of these systems in different emergencies of SLOs, focusing on a medium job load for simplicity. As shown in Figures~\ref{fig:e2e-slo} and~\ref{fig:e2e-slo-cost}, \SysName{} consistently outperforms baseline systems with at least 10\% SLO violation reduction across varying SLO levels. When the SLO emergence is set as 0.5, more LPT jobs are executed on multiple GPUs. Thus, INFless is more likely to suffer from the long waiting delay incurred by the instance initialization, as discussed in~\ref{sub:dl-inference-limitations}. Hence, INFless even achieves very high SLO violation as ElasticFlow. In terms of resource cost, compared to INFless, \SysName{} reduces the expenses by 38\%, 23\%, and 17\% at SLO levels \(S =\) 0.5, 1.0, and 1.5, respectively. Compared to ElasticFlow, the cost savings of \SysName{} are even more pronounced: up to 70\% at \(S =\) 1.5. In summary, \SysName{} stands out for its superior performance in both SLO violation reduction and cost efficiency. 

\begin{table}[t]\centering
\caption{\small Heavy Workload Evaluation.}\label{tab:simulator}
\vspace{-10pt}
\resizebox{1.0\linewidth}{!}{
\begin{tabular}{lrcccc}\toprule
\textbf{Heavy Setting} &\textbf{Metric} &\textbf{PromptTuner} &\textbf{INFless} &\textbf{ElasticFlow} \\\midrule
\multirow{2}{*}{LLaMA-30B} &SLO Violation $\downarrow$(\%) &28.4 &38.9 &82.3 \\
&Cost $\downarrow$ (\$) &38.8 &46.4 &69.4 \\\midrule
\multirow{2}{*}{Qwen7B-R1} &SLO Violation $\downarrow$(\%) &23.1 &36.2 &74.9 \\
                                &Cost $\downarrow$ (\$) &30.7 &42.8 &70.1 \\\midrule
\multirow{2}{*}{Large-Scale} &SLO Violation $\downarrow$ (\%) &25.4 &57.1 &78.2 \\
&Cost $\downarrow$ (\$) &57.2 &65.9 &99.1 \\
\bottomrule
\end{tabular}}
\vspace{-15pt}
\end{table}

\noindent\textbf{LLaMA-30B Evaluation.} A single replica of LLaMA-30B is hosted across four GPUs, with tensor parallelism employed to facilitate prompt tuning. Due to limited GPU availability, the experiment for LLaMA-30B is conducted separately. Table~\ref{tab:simulator} compares the SLO violation rates and resource costs of \SysName{} with two baseline methods. \SysName{} reduces the SLO violation rate by 1.36-2.90\(\times \) and resource costs by 1.20-1.79\( \times \) compared to another two baselines. These results suggest that \SysName{} sustains its superior performance when managing heavy LLM workloads.

\noindent\textbf{Qwen7B-R1 Evaluation.} The maximum sequence length for Qwen7B-R1 is 32,768 tokens. To accommodate this, we utilize four GPUs to host a replica of Qwen7B-R1 using tensor parallelism. Furthermore, we employ a cluster of 32 GPUs to evaluate the performance of Qwen7B-R1. Table~\ref{tab:simulator} compares the SLO violation rates and resource costs of \SysName{} with two baseline methods. \SysName{} reduces the SLO violation rate by 1.56-3.24\(\times \) and resource costs by 1.39-2.28\( \times \) compared to another two baselines. These results suggest that \SysName{} sustains its superior performance when handling long-sequence samples.

\noindent\textbf{Scalability Evaluation.} We measure the performance of \SysName{} in large-scale GPU clusters. Because of limited available GPUs, we perform one experiment for each system on a cluster of up to 96 GPUs. We increase Figure~\ref{fig:e2e-slo}'s medium job loads proportionally to match the maximal amount of provisioned GPUs. Table~\ref{tab:simulator} compares the SLO violation and resource costs of \SysName{} with the other two baselines. The performance gain of \SysName{} over other baselines is enlarged with the increase of provisioned GPUs. With more workloads and GPUs, \SysName{} can exploit dynamic resource allocation to obtain better scheduling decisions. Additionally, the average/maximal scheduling overhead is  13/67 ms, making it not a performance bottleneck in \SysName{}. The small scheduling overhead strengthens our belief that \SysName{} can attain satisfactory performance in a large-scale GPU cluster.

\subsection{Evaluation of Key Components}
\label{sec:eval-ablation-studies}

\noindent\textbf{Prompt \& Runtime Reusing.} Figures~\ref{fig:prompt-sharing-slo} and ~\ref{fig:runtime-sharing-slo} show the benefits of \textit{prompt reusing} (P.R.) and \textit{runtime reusing} (R.R.) to SLO guarantee and cost-effectiveness over different SLO levels. First, prompt reusing can reduce SLO violations by 13-23\% and cost savings by 30-40\%. For the stringent SLO, the Prompt Bank (i.e., prompt reusing) saves GPU time by satisfying more SLOs of LPT jobs. In relaxed SLO scenarios, the Prompt Bank can reduce the number of GPUs allocated to warm GPU pools. \SysName{} particularly benefits from \textit{runtime reusing} by mitigating the GPU allocation overhead, enhancing SLO attainment. However, the cost savings from \textit{runtime reusing} are not comparable to that of \textit{prompt reusing}.

\begin{figure}
\begin{subfigure}{.24\textwidth}
  \centering
  \includegraphics[width=\textwidth]{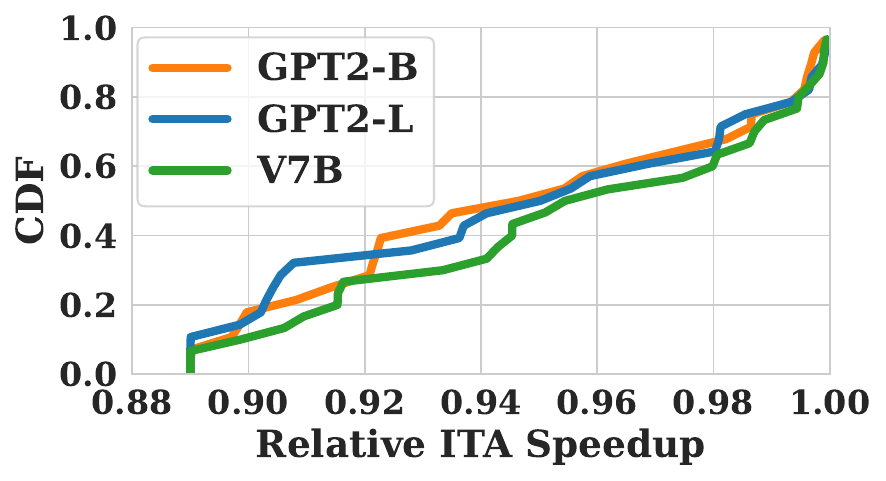}
  \caption{Score versus Ideal.}
  \label{fig:evaluator-ideal}
\end{subfigure}
\begin{subfigure}{.23\textwidth}
  \centering
  \includegraphics[width=\textwidth]{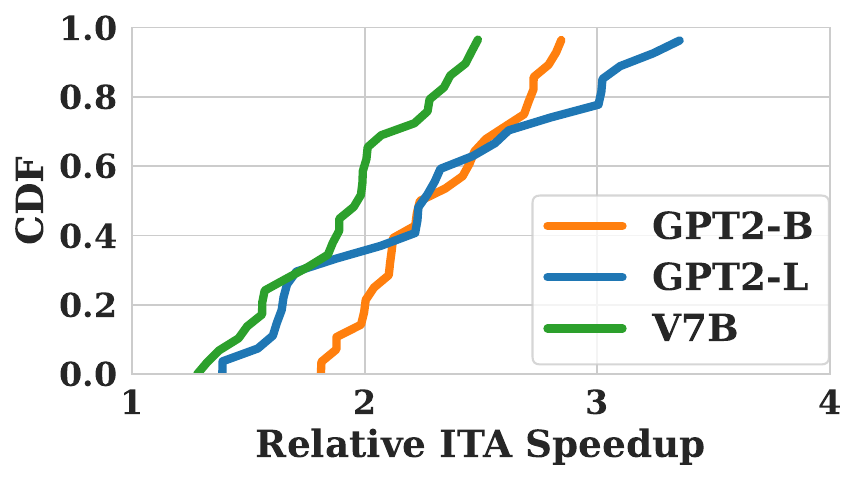}
  \caption{Score versus Induction.}
  \label{fig:evaluator-baseline}
\end{subfigure}
\vspace{-20pt}
\caption{\small Distributions of relative ITA speedup of score candidate to (a) ideal candidate; (b) induction candidate.}
\label{fig:evaluator}
\vspace{-12pt}
\end{figure}

\begin{figure} 
\vspace{-5pt}
\begin{subfigure}{.22\textwidth}
  \centering
  \includegraphics[width=\textwidth]{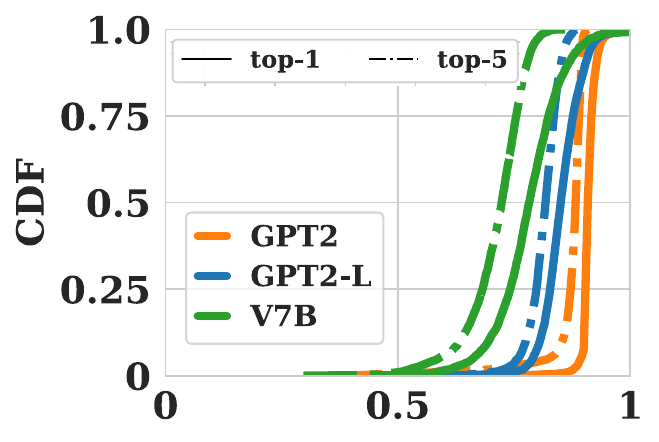}
  \caption{Prompt Similarity.}
  \label{fig:promt-sim}
\end{subfigure}
\begin{subfigure}{.24\textwidth}
  \centering
  \includegraphics[width=\textwidth]{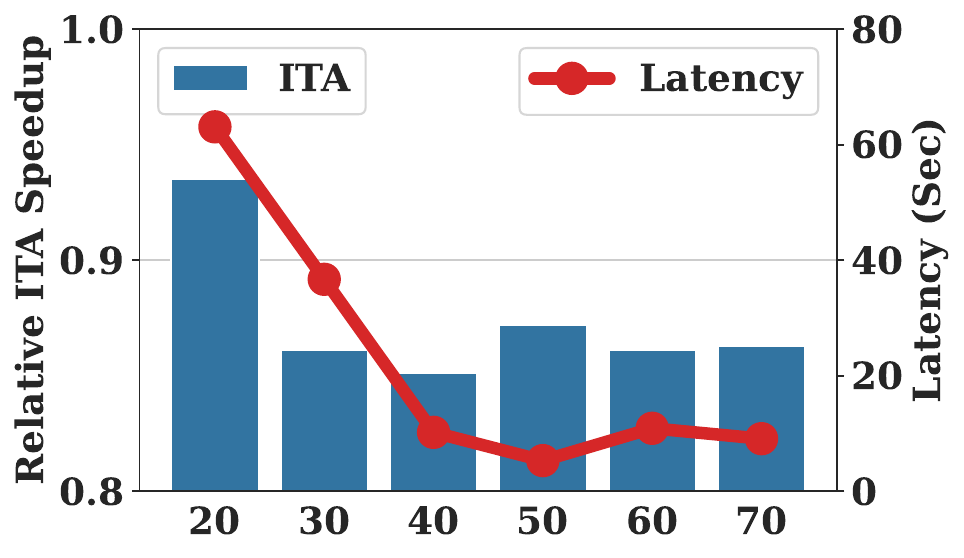}
  \caption{Varying $K$.}
  \label{fig:bank-latency}
\end{subfigure}
\vspace{-10pt}
\caption{\small Performance of the \emph{two-layer structure}: (a) Distribution of prompt similarity; (b) latency and average relative TTA of varying numbers of groups.}
\label{fig:bank}
\vspace{-15pt}
\end{figure}

\noindent\textbf{Impact of Allocation from Warm GPU pool.} The GPU allocation from a warm GPU pool enables simultaneous multi-GPU allocation, effectively mitigating the initialization overhead associated with multi-GPU instances, as discussed in \S~\ref{sub:dl-inference-limitations}. We implement a baseline policy (without the warm allocator) that immediately allocates warm GPUs to each function instance (\S\ref{sec:preemptive-elastic-execution}), disregarding the constraints of simultaneous allocation within the same LPT request. The second and third columns of Table~\ref{tab:impact-of-workload-scheduler} demonstrate that our proposed allocation policy reduces the SLO violation rate by a factor of 2.24, with only a modest increase in cost at an SLO level of 1.0 and a medium job load. These results suggest that the warm GPU allocator has a positive impact on SLO attainment. 

\noindent\textbf{Impact of DelaySchedulable Function.} The DelaySchedulable function delays the execution of certain LPT requests to fully utilize GPUs from the warm GPU pool, reducing the GPU amount in a warm GPU pool while minimizing the SLO violation rate. As shown in the second and fourth columns of Table~\ref{tab:impact-of-workload-scheduler}, this function reduces the SLO violation rate and resource costs by 1.27\(\times\) and 1.16\(\times\), respectively, at an SLO level of 1.0 and a medium job load. This empirically demonstrates the effectiveness of the DelaySchedulable function in optimizing both SLO violations and resource costs. 

\noindent\textbf{Impact of Latency Budget.} To evaluate the effectiveness of the latency budget, we establish a baseline in which the Prompt Bank is triggered for each incoming request. As shown in Table~\ref{tab:impact-of-workload-scheduler}, the latency budget leads to a reduction in both SLO violations and costs by 1.31\(\times\) and 1.02\(\times\) respectively. The latency budget proves to be a beneficial operation in \SysName{}.


\noindent\textbf{Window Size of Allocation from Cold GPU Pool.}  We investigate how the window size of the cold GPU allocator affects the performance of \SysName{}. A smaller window size causes GPUs to be removed from the warm GPU pool frequently, increasing the SLO violation. A larger window size may make \SysName{} less responsive to traffic, increasing resource costs. Figure~\ref{fig:window-time} presents various window sizes and shows that setting 60 seconds strikes a satisfactory balance between the SLO violation and cost.

\noindent\textbf{Varying Prompt Bank Size.} Due to the heavy evaluation costs of Prompt Bank, we only choose GPT2-Base, GPT2-Large, and Vicuna-7B to investigate. We analyze the impact of the number of prompt candidates on the scheduling performance of \SysName{}. We set the maximum size as 3,000 due to the limited number of free-of-use high-quality prompts. Figure~\ref{fig:vary-prompt-bank-size} depicts that the SLO violation and cost vary over different sizes of the Prompt Bank. A larger Prompt Bank incurs larger execution overhead, while a smaller one may reduce the potential speedup benefits derived from effective initial prompts. When the size drops to 2,000, both SLO violations and costs increase significantly, highlighting the importance of maintaining prompt diversity of the Prompt Bank.

\begin{table}[t]\centering
\caption{\small Impact of key components in Workload Scheduler.}\label{tab:impact-of-workload-scheduler}
\vspace{-10pt}
\resizebox{0.98\linewidth}{!}{
\begin{tabular}{lccccc}\toprule
\textbf{Metric} & \makecell{\textbf{Workload} \\ \textbf{Scheduler}} & \makecell{\textbf{w/o} \\ \textbf{Warm Allocator}} & \makecell{\textbf{w/o} \\ \textbf{DelaySchedulable}} & \makecell{\textbf{w/o} \\ \textbf{Latency Budget}} \\\midrule
SLO Violation $\downarrow$ (\%) &12.4 &27.8 &15.6 & 16.3 \\
Cost $\downarrow$ (\$) &22.9 &20.9 &26.6 & 23.2 \\
\bottomrule
\end{tabular}}
\vspace{-15pt}
\end{table}


\noindent\textbf{Score Metric.}
We term \textit{score candidate},  \textit{ideal candidate}, and \textit{induction candidate} as the prompts selected by proposed metric (Eqn.~\ref{eq:score}), ideal baseline, and induction baseline, respectively. Figure~\ref{fig:evaluator-ideal} shows the distributions of relative ITA performance between the score candidate and ideal candidate from 120 LPT tasks of three LLMs. The ITA performance of most score candidates exceeds \( 90\% \) of that of ideal candidates. Figure~\ref{fig:evaluator-baseline} presents the distributions of relative ITA performance between the score candidates and induction candidates. The score candidates outperform the induction candidates and yield at least 1.81\( \times \), 1.38\( \times \), 1.28\( \times \) ITA speedup for GPT-Base, GPT-Large, and Vicuna-7B, respectively. GPT-B achieves the highest ITA speedup benefits (1.8–2.8\(\times\)), as its generality is weak compared to Vicuna-7B, leading to less effective initial prompt generation by itself. Conversely, V-7B achieves a minimum ITA speedup of 1.28\( \times \) compared to induction candidates. This analysis highlights that our \(\mathtt{score}\) method can identify near-optimal initial prompts, delivering superior ITA performance over induction initialization across various tasks and LLMs.

\noindent\textbf{Two-layer Data Structure.} Figure~\ref{fig:promt-sim} shows the CDF of top-1 (solid line) and top-5 (dashed line) cosine similarity of the activation features in our curated prompt candidate set across varying LLMs. This high similarity motivates us to design a two-level data structure to group similar prompt candidates. Furthermore, we verify whether clustering similar prompt candidates degrades the ITA performance of the identified initial prompt and reduces the selection latency. We fix the number of evaluation samples to 16 and the LLM to GPT2-Base. Figure~\ref{fig:bank-latency} shows the impact of the cluster counts on the relative ITA speedup compared to the ideal candidate and the average selection latency. Using more groups does not cause considerable ITA performance loss. For GPT2-Large and Vicuna-7B, the impact of cluster counts on ITA speedup presents a similar trend. Also, we are concerned about the latency overhead and set the number of clusters as 50 for \SysName{}. Then the average latency is 5.3 seconds for GPT2-Base, 6.1 seconds for GPT2-Large, and 9.2 seconds for Vicuna-7B, respectively. As a reference, it takes approximately 2.5, 2.9, and 4.5 hours for GPT-Base, GPT-Large, and Vicuna-7B, respectively, when $K$ is set to 1. Overall, the two-layer data structure efficiently balances speedup benefits with the overhead of initial prompt selection.

\section{Related Works}
\label{sec:related-works}
\noindent\textbf{Workload Scheduling.} Substantial schedulers are designed to satisfy the latency SLOs. Training systems~\cite{Chronus,GENIE,ASTRAEA,UniSched,autosched,ymir,titan,gao2025icefrog} consider resource elasticity to adjust the GPU allocation. Inference systems~\cite{INFaaS,INFless,cui2021abacus,Arpan2020clockwork,FaaSwap} exploit GPU sharing and request batching to meet SLOs while improving the GPU utilization. Our system benefits from their designs, including resource elasticity and runtime reusing.

\noindent\textbf{LLM Systems.} 
The significance of LLMs attracts researchers to design specialized systems to support their execution. Many LLM systems~\cite{FlexFlow,zheng2022alpa,MegatronLM,narayanan2019pipedream,huang2019gpipe,chen2025sppo,guo2025adaptis} focus on automatic discovery of parallelism strategies for deploying LLM training on thousands of GPUs. In our scenario, LPT has significantly smaller communication overhead and GPU requests (at most tens) than LLM training. Thus, these strategies are not well-suited to LPT. Many LLM inference works~\cite{yu2022orca,kwon2023efficient,S-LORA,lightllm,chen2023punica,fu2024serverlessllm,gao2025rethinkingkvcache} mainly address the mismatch between the computation-bound prefill phase and the memory-bound decoding phase, along with the heavy KV cache. However, since many prompt tuning algorithms do not involve the decoding phase and KV cache management, \SysName{} cannot directly adopt solutions proposed by these inference systems. Instead, \SysName{} utilizes prompt sharing to accelerate LPT.





\noindent\textbf{Parameter-Efficient Fine-Tuning.}
Recent works design various parameter-efficient fine-tuning methods for LLMs, including LoRA~\cite{hu2021lora}, Prefix-tuning~\cite{li2021prefix}, P-Tuning~\cite{liu2021p}, and Prompt tuning~\cite{liu2023gptunderstand,lester2021power}. Prefix-tuning and P-tuning are extensions of prompt tuning, and \SysName{} can treat them as LPT workloads to determine the GPU allocation. The LLM community has also introduced other methods like gradient-based approaches~\cite{li2021prefix,TEXTBOX}, zero order~\cite{malladi2023fine}, and reinforcement learning~\cite{deng2022rlprompt} to optimize the prompts. 

\section{Conclusion}
This paper presents \SysName{}, an SLO-aware elastic system for managing LPT workloads. We take advantage of \textit{prompt reusing} to develop the Prompt Bank for expediting LPT workloads. We also exploit the \textit{runtime reusing} to reduce the GPU allocation overhead for resource elasticity. Our extensive experiments demonstrate the superiority of \SysName{} in SLO attainment and cost reduction.


\bibliographystyle{plain}
\bibliography{references}

@STRING{mlsys	= "Proceedings of Machine Learning and Systems" }

@STRING{sc	= "Proceedings of the International Conference for High
		  Performance Computing, Networking, Storage and Analysis" }

@inproceedings{Arpan2020clockwork,
  title     = {Serving {DNNs} like Clockwork: Performance Predictability
               from the Bottom Up},
  author    = {Arpan Gujarati and Reza Karimi and Safya Alzayat and Wei
               Hao and Antoine Kaufmann and Ymir Vigfusson and Jonathan
               Mace},
  booktitle = {14th USENIX Symposium on Operating Systems Design and
               Implementation (OSDI 20)},
  year      = {2020}
}

@article{ASTRAEA,
  title   = {ASTRAEA: A Fair Deep Learning Scheduler for Multi-tenant
             GPU Clusters},
  author  = {Ye, Zhisheng and Sun, Peng and Gao, Wei and Zhang, Tianwei
             and Wang, Xiaolin and Yan, Shengen and Luo, Yingwei},
  journal = {IEEE Transactions on Parallel and Distributed Systems},
  year    = {2021}
}

@inproceedings{Chronus,
  title     = {Chronus: A Novel Deadline-aware Scheduler for Deep
               Learning Training Jobs},
  author    = {Gao, Wei and Ye, Zhisheng and Sun, Peng and Wen, Yonggang
               and Zhang, Tianwei},
  booktitle = {Proceedings of the ACM Symposium on Cloud Computing},
  year      = {2021},
  series    = {SoCC '21}
}

@inproceedings{cui2021abacus,
  title     = {Enable Simultaneous DNN Services Based on Deterministic
               Operator Overlap and Precise Latency Prediction},
  author    = {Cui, Weihao and Zhao, Han and Chen, Quan and Zheng,
               Ningxin and Leng, Jingwen and Zhao, Jieru and Song, Zhuo
               and Ma, Tao and Yang, Yong and Li, Chao and Guo, Minyi},
  booktitle = {Proceedings of the International Conference for High
               Performance Computing, Networking, Storage and Analysis},
  year      = {2021},
  series    = {SC '21}
}

@inproceedings{FlexFlow,
  title     = {Beyond Data and Model Parallelism for Deep Neural
               Networks},
  author    = {Zhihao Jia and Matei Zaharia and Alex Aiken},
  booktitle = {Proceedings of Machine Learning and Systems},
  year      = {2019},
  series    = {MLSys '19}
}

@article{GENIE,
  title   = {Deep Learning Research and Development Platform:
             Characterizing and Scheduling with QoS Guarantees on GPU
             Clusters},
  author  = {Chen, Zhaoyun and Quan, Wei and Wen, Mei and Fang, Jianbin
             and Yu, Jie and Zhang, Chunyuan and Luo, Lei},
  journal = {IEEE Transactions on Parallel and Distributed Systems},
  year    = {2020}
}

@inproceedings{GPT-3,
  title     = {Language Models are Few-Shot Learners},
  author    = {Brown, Tom and Mann, Benjamin and Ryder, Nick and Subbiah,
               Melanie and Kaplan, Jared D and Dhariwal, Prafulla and
               Neelakantan, Arvind and Shyam, Pranav and Sastry, Girish
               and Askell, Amanda and Agarwal, Sandhini and Herbert-Voss,
               Ariel and Krueger, Gretchen and Henighan, Tom and Child,
               Rewon and Ramesh, Aditya and Ziegler, Daniel and Wu,
               Jeffrey and Winter, Clemens and Hesse, Chris and Chen, Mark
               and Sigler, Eric and Litwin, Mateusz and Gray, Scott and
               Chess, Benjamin and Clark, Jack and Berner, Christopher and
               McCandlish, Sam and Radford, Alec and Sutskever, Ilya and
               Amodei, Dario},
  booktitle = {Advances in Neural Information Processing Systems},
  year      = {2020},
  series    = {NeurIPS '20}
}

@inproceedings{Helios,
  title     = {Characterization and Prediction of Deep Learning Workloads
               in Large-Scale GPU Datacenters},
  author    = {Hu, Qinghao and Sun, Peng and Yan, Shengen and Wen,
               Yonggang and Zhang, Tianwei},
  booktitle = {Proceedings of the International Conference for High
               Performance Computing, Networking, Storage and Analysis},
  year      = {2021},
  series    = {SC '21}
}

@inproceedings{INFaaS,
  title     = {INFaaS: Automated Model-less Inference Serving},
  author    = {Francisco Romero and Qian Li and Neeraja J. Yadwadkar and
               Christos Kozyrakis},
  booktitle = {2021 {USENIX} Annual Technical Conference},
  year      = {2021},
  series    = {{USENIX} {ATC} '21}
}

@inproceedings{MegatronLM,
  title     = {Efficient large-scale language model training on GPU
               clusters using megatron-LM},
  author    = {Narayanan, Deepak and Shoeybi, Mohammad and Casper, Jared
               and LeGresley, Patrick and Patwary, Mostofa and
               Korthikanti, Vijay and Vainbrand, Dmitri and Kashinkunti,
               Prethvi and Bernauer, Julie and Catanzaro, Bryan and
               Phanishayee, Amar and Zaharia, Matei},
  booktitle = {Proceedings of the International Conference for High
               Performance Computing, Networking, Storage and Analysis},
  year      = {2021},
  series    = {SC '21}
}

@misc{Memcached,
  title        = {Memcached},
  year         = 2025,
  howpublished = {\url{https://memcached.org/}}
}

@inproceedings{MLaaS,
  title     = {{MLaaS} in the Wild: Workload Analysis and Scheduling in
               {Large-Scale} Heterogeneous {GPU} Clusters},
  author    = {Weng, Qizhen and Xiao, Wencong and Yu, Yinghao and Wang,
               Wei and Wang, Cheng and He, Jian and Li, Yong and Zhang,
               Liping and Lin, Wei and Ding, Yu},
  booktitle = {19th USENIX Symposium on Networked Systems Design and
               Implementation},
  year      = {2022},
  series    = {NSDI '22}
}

@inproceedings{Philly,
  title     = {Analysis of Large-Scale Multi-Tenant {GPU} Clusters for
               {DNN} Training Workloads},
  author    = {Myeongjae Jeon and Shivaram Venkataraman and Amar
               Phanishayee and Junjie Qian and Wencong Xiao and Fan Yang},
  booktitle = {2019 {USENIX} Annual Technical Conference},
  year      = {2019},
  series    = {{USENIX} {ATC} '19}
}

@inproceedings{ServerlessATC20,
  title     = {Serverless in the Wild: Characterizing and Optimizing the
               Serverless Workload at a Large Cloud Provider},
  author    = {Mohammad Shahrad and Rodrigo Fonseca and Inigo Goiri and
               Gohar Chaudhry and Paul Batum and Jason Cooke and Eduardo
               Laureano and Colby Tresness and Mark Russinovich and
               Ricardo Bianchini},
  booktitle = {2020 {USENIX} Annual Technical Conference},
  year      = {2020},
  series    = {{USENIX} {ATC} '20}
}

@inproceedings{zhang2019mark,
  title     = {{MArk}: Exploiting Cloud Services for {Cost-Effective},
               {SLO-Aware} Machine Learning Inference Serving},
  author    = {Chengliang Zhang and Minchen Yu and Wei Wang and Feng
               Yan},
  booktitle = {2019 USENIX Annual Technical Conference (USENIX ATC 19)},
  year      = {2019}
}

@article{GPT-2,
  title={Language models are unsupervised multitask learners},
  author={Radford, Alec and Wu, Jeffrey and Child, Rewon and Luan, David and Amodei, Dario and Sutskever, Ilya and others},
  journal={OpenAI blog},
  volume={1},
  number={8},
  pages={9},
  year={2019}
}

@article{lester2021power,
  title={The power of scale for parameter-efficient prompt tuning},
  author={Lester, Brian and Al-Rfou, Rami and Constant, Noah},
  journal={arXiv preprint arXiv:2104.08691},
  year={2021}
}

@inproceedings{ModelKeeper,
  title={ModelKeeper: Accelerating DNN Training via Automated Training Warmup},
  author={Fan Lai and Yinwei Dai and Harsha V. Madhyastha and Mosharaf Chowdhury},
  booktitle={USENIX Symposium on Networked Systems Design and Implementation (NSDI)},
  year={2023}
}

@article{tripuraneni2020theory,
  title={On the theory of transfer learning: The importance of task diversity},
  author={Tripuraneni, Nilesh and Jordan, Michael and Jin, Chi},
  journal={Advances in Neural Information Processing Systems},
  volume={33},
  pages={7852--7862},
  year={2020}
}

@article{NLPTransfer,
  title={Exploring and predicting transferability across NLP tasks},
  author={Vu, Tu and Wang, Tong and Munkhdalai, Tsendsuren and Sordoni, Alessandro and Trischler, Adam and Mattarella-Micke, Andrew and Maji, Subhransu and Iyyer, Mohit},
  journal={arXiv preprint arXiv:2005.00770},
  year={2020}
}

@article{hu2021lora,
  title={Lora: Low-rank adaptation of large language models},
  author={Hu, Edward J and Shen, Yelong and Wallis, Phillip and Allen-Zhu, Zeyuan and Li, Yuanzhi and Wang, Shean and Wang, Lu and Chen, Weizhu},
  journal={arXiv preprint arXiv:2106.09685},
  year={2021}
}

@article{zheng2022alpa,
  title={Alpa: Automating Inter-and Intra-Operator Parallelism for Distributed Deep Learning},
  author={Zheng, Lianmin and Li, Zhuohan and Zhang, Hao and Zhuang, Yonghao and Chen, Zhifeng and Huang, Yanping and Wang, Yida and Xu, Yuanzhong and Zhuo, Danyang and Gonzalez, Joseph E and others},
  journal={arXiv preprint arXiv:2201.12023},
  year={2022}
}

@misc{vicuna,
    title = {Vicuna: An Open-Source Chatbot Impressing GPT-4 with 90\%* ChatGPT Quality},
    url = {https://lmsys.org/blog/2023-03-30-vicuna/},
    author = {Chiang, Wei-Lin and Li, Zhuohan and Lin, Zi and Sheng, Ying and Wu, Zhanghao and Zhang, Hao and Zheng, Lianmin and Zhuang, Siyuan and Zhuang, Yonghao and Gonzalez, Joseph E. and Stoica, Ion and Xing, Eric P.},
    month = {March},
    year = {2023}
}

@article{llama-7b,
  title={LLaMA: open and efficient foundation language models, 2023},
  author={Touvron, Hugo and Lavril, Thibaut and Izacard, Gautier and Martinet, Xavier and Lachaux, Marie-Anne and Lacroix, Timoth{\'e}e and Rozi{\`e}re, Baptiste and Goyal, Naman and Hambro, Eric and Azhar, Faisal and others},
  journal={URL https://arxiv. org/abs/2302.13971}
}

@misc{chatgpt,
      title={GPT-4 Technical Report}, 
      author={OpenAI},
      year={2023},
      eprint={2303.08774},
      archivePrefix={arXiv},
      primaryClass={cs.CL}
}

@inproceedings{ROC,
    title = "A Corpus and Cloze Evaluation for Deeper Understanding of Commonsense Stories",
    author = "Mostafazadeh, Nasrin  and
      Chambers, Nathanael  and
      He, Xiaodong  and
      Parikh, Devi  and
      Batra, Dhruv  and
      Vanderwende, Lucy  and
      Kohli, Pushmeet  and
      Allen, James",
    booktitle = "Proceedings of the 2016 Conference of the North {A}merican Chapter of the Association for Computational Linguistics: Human Language Technologies",
    month = jun,
    year = "2016",
    address = "San Diego, California",
    publisher = "Association for Computational Linguistics",
    url = "https://aclanthology.org/N16-1098",
    doi = "10.18653/v1/N16-1098",
    pages = "839--849",
}

@inproceedings{WP,
    title = "Hierarchical Neural Story Generation",
    author = "Fan, Angela  and
      Lewis, Mike  and
      Dauphin, Yann",
    booktitle = "Proceedings of the 56th Annual Meeting of the Association for Computational Linguistics (Volume 1: Long Papers)",
    month = jul,
    year = "2018",
    address = "Melbourne, Australia",
    publisher = "Association for Computational Linguistics",
    url = "https://aclanthology.org/P18-1082",
    doi = "10.18653/v1/P18-1082",
    pages = "889--898",
}

@inproceedings{TEXTBOX,
    title = "{T}ext{B}ox: A Unified, Modularized, and Extensible Framework for Text Generation",
    author = "Li, Junyi  and  Tang, Tianyi  and  He, Gaole  and  Jiang, Jinhao  and  Hu, Xiaoxuan  and  Xie, Puzhao  and  Chen, Zhipeng  and  Yu, Zhuohao  and  Zhao, Wayne Xin  and  Wen, Ji-Rong",
    booktitle = "Proceedings of the 59th Annual Meeting of the Association for Computational Linguistics and the 11th International Joint Conference on Natural Language Processing: System Demonstrations",
    month = aug,
    year = "2021",
    address = "Online",
    publisher = "Association for Computational Linguistics",
    url = "https://aclanthology.org/2021.acl-demo.4",
    doi = "10.18653/v1/2021.acl-demo.4",
    pages = "30--39",
}

@inproceedings{PC,
    title = "Personalizing Dialogue Agents: {I} have a dog, do you have pets too?",
    author = "Zhang, Saizheng  and
      Dinan, Emily  and
      Urbanek, Jack  and
      Szlam, Arthur  and
      Kiela, Douwe  and
      Weston, Jason",
    booktitle = "Proceedings of the 56th Annual Meeting of the Association for Computational Linguistics (Volume 1: Long Papers)",
    month = jul,
    year = "2018",
    address = "Melbourne, Australia",
    publisher = "Association for Computational Linguistics",
    url = "https://aclanthology.org/P18-1205",
    doi = "10.18653/v1/P18-1205",
    pages = "2204--2213",
}

@article{COQAQG,
    title = "{C}o{QA}: A Conversational Question Answering Challenge",
    author = "Reddy, Siva  and
      Chen, Danqi  and
      Manning, Christopher D.",
    journal = "Transactions of the Association for Computational Linguistics",
    volume = "7",
    year = "2019",
    address = "Cambridge, MA",
    publisher = "MIT Press",
    url = "https://aclanthology.org/Q19-1016",
    pages = "249--266",
}

@inproceedings{SAMSUM,
    title = "{SAMS}um Corpus: A Human-annotated Dialogue Dataset for Abstractive Summarization",
    author = "Gliwa, Bogdan  and
      Mochol, Iwona  and
      Biesek, Maciej  and
      Wawer, Aleksander",
    booktitle = "Proceedings of the 2nd Workshop on New Frontiers in Summarization",
    month = nov,
    year = "2019",
    address = "Hong Kong, China",
    publisher = "Association for Computational Linguistics",
    url = "https://aclanthology.org/D19-5409",
    doi = "10.18653/v1/D19-5409",
    pages = "70--79",
}

@misc{WIKIP,
  title = {WikiPlots},
  howpublished = {\url{https://github.com/markriedl/WikiPlots}},
  year = {2025},
}

@inproceedings{CMV,
    title = "{PAIR}: Planning and Iterative Refinement in Pre-trained Transformers for Long Text Generation",
    author = "Hua, Xinyu  and
      Wang, Lu",
    booktitle = "Proceedings of the 2020 Conference on Empirical Methods in Natural Language Processing (EMNLP)",
    month = nov,
    year = "2020",
    address = "Online",
    publisher = "Association for Computational Linguistics",
    url = "https://aclanthology.org/2020.emnlp-main.57",
    doi = "10.18653/v1/2020.emnlp-main.57",
    pages = "781--793",
}

@article{QUORA,
    title = "Syntax-Guided Controlled Generation of Paraphrases",
    author = "Kumar, Ashutosh  and
      Ahuja, Kabir  and
      Vadapalli, Raghuram  and
      Talukdar, Partha",
    journal = "Transactions of the Association for Computational Linguistics",
    volume = "8",
    year = "2020",
    address = "Cambridge, MA",
    publisher = "MIT Press",
    url = "https://aclanthology.org/2020.tacl-1.22",
    pages = "329--345",
}

@misc{promptperfect,
title = {PromptPerfect},
howpublished = {\url{https://promptperfect.jina.ai/home}}
}

@misc{prompthero,
title = {PromptHero},
howpublished = {\url{https://prompthero.com/}}
}

@misc{mergeflow,
title = {mergeflow},
howpublished = {\url{https://mergeflow.com/}}
}

@misc{midjourney,
title = {Midjourney},
howpublished = {\url{https://www.midjourney.com}}
}

@inproceedings{INFless,
  title={INFless: a native serverless system for low-latency, high-throughput inference},
  author={Yang, Yanan and Zhao, Laiping and Li, Yiming and Zhang, Huanyu and Li, Jie and Zhao, Mingyang and Chen, Xingzhen and Li, Keqiu},
  booktitle={Proceedings of the 27th ACM International Conference on Architectural Support for Programming Languages and Operating Systems},
  pages={768--781},
  year={2022}
}

@inproceedings{elasticflow,
  title={ElasticFlow: An Elastic Serverless Training Platform for Distributed Deep Learning},
  author={Gu, Diandian and Zhao, Yihao and Zhong, Yinmin and Xiong, Yifan and Han, Zhenhua and Cheng, Peng and Yang, Fan and Huang, Gang and Jin, Xin and Liu, Xuanzhe},
  booktitle={Proceedings of the 28th ACM International Conference on Architectural Support for Programming Languages and Operating Systems, Volume 2},
  pages={266--280},
  year={2023}
}

@article{yao2023tree,
  title={Tree of thoughts: Deliberate problem solving with large language models},
  author={Yao, Shunyu and Yu, Dian and Zhao, Jeffrey and Shafran, Izhak and Griffiths, Thomas L and Cao, Yuan and Narasimhan, Karthik},
  journal={arXiv preprint arXiv:2305.10601},
  year={2023}
}

@article{FaaSwap,
  title={FaaSwap: SLO-Aware, GPU-Efficient Serverless Inference via Model Swapping},
  author={Yu, Minchen and Wang, Ao and Chen, Dong and Yu, Haoxuan and Luo, Xiaonan and Li, Zhuohao and Wang, Wei and Chen, Ruichuan and Nie, Dapeng and Yang, Haoran},
  journal={arXiv preprint arXiv:2306.03622},
  year={2023}
}

@inproceedings{su-etal-2022-transferability,
    title = "On Transferability of Prompt Tuning for Natural Language Processing",
    author = "Su, Yusheng  and
      Wang, Xiaozhi  and
      Qin, Yujia  and
      Chan, Chi-Min  and
      Lin, Yankai  and
      Wang, Huadong  and
      Wen, Kaiyue  and
      Liu, Zhiyuan  and
      Li, Peng  and
      Li, Juanzi  and
      Hou, Lei  and
      Sun, Maosong  and
      Zhou, Jie",
    booktitle = "Proceedings of the 2022 Conference of the North American Chapter of the Association for Computational Linguistics: Human Language Technologies",
    month = jul,
    year = "2022",
    address = "Seattle, United States",
    publisher = "Association for Computational Linguistics",
    url = "https://aclanthology.org/2022.naacl-main.290",
    doi = "10.18653/v1/2022.naacl-main.290",
    pages = "3949--3969"
}

@inproceedings{SIREN,
  title={Distributed machine learning with a serverless architecture},
  author={Wang, Hao and Niu, Di and Li, Baochun},
  booktitle={IEEE INFOCOM},
  pages={1288--1296},
  year={2019}
}

@inproceedings{lambdaml,
  title={Towards demystifying serverless machine learning training},
  author={Jiang, Jiawei and Gan, Shaoduo and Liu, Yue and Wang, Fanlin and Alonso, Gustavo and Klimovic, Ana and Singla, Ankit and Wu, Wentao and Zhang, Ce},
  booktitle={Proceedings of the 2021 International Conference on Management of Data},
  pages={857--871},
  year={2021}
}

@misc{awesome_chatgpt_prompts,
  author = {Fatih Erikli},
  title = {Awesome ChatGPT Prompts},
  year = {2022},
  publisher = {GitHub},
  howpublished = {\url{https://github.com/f/awesome-chatgpt-prompts/}},
}

@inproceedings{li2021prefix,
  title={Prefix-Tuning: Optimizing Continuous Prompts for Generation},
  author={Li, Xiang Lisa and Liang, Percy},
  booktitle={Proceedings of the 59th Annual Meeting of the Association for Computational Linguistics and the 11th International Joint Conference on Natural Language Processing (Volume 1: Long Papers)},
  pages={4582--4597},
  year={2021}
}

@article{malladi2023fine,
  title={Fine-Tuning Language Models with Just Forward Passes},
  author={Malladi, Sadhika and Gao, Tianyu and Nichani, Eshaan and Damian, Alex and Lee, Jason D and Chen, Danqi and Arora, Sanjeev},
  journal={arXiv preprint arXiv:2305.17333},
  year={2023}
}

@article{deng2022rlprompt,
  title={Rlprompt: Optimizing discrete text prompts with reinforcement learning},
  author={Deng, Mingkai and Wang, Jianyu and Hsieh, Cheng-Ping and Wang, Yihan and Guo, Han and Shu, Tianmin and Song, Meng and Xing, Eric P and Hu, Zhiting},
  journal={arXiv preprint arXiv:2205.12548},
  year={2022}
}

@article{besta2023graph,
  title={Graph of thoughts: Solving elaborate problems with large language models},
  author={Besta, Maciej and Blach, Nils and Kubicek, Ales and Gerstenberger, Robert and Gianinazzi, Lukas and Gajda, Joanna and Lehmann, Tomasz and Podstawski, Michal and Niewiadomski, Hubert and Nyczyk, Piotr and others},
  journal={arXiv preprint arXiv:2308.09687},
  year={2023}
}

@article{zhou2022large,
  title={Large language models are human-level prompt engineers},
  author={Zhou, Yongchao and Muresanu, Andrei Ioan and Han, Ziwen and Paster, Keiran and Pitis, Silviu and Chan, Harris and Ba, Jimmy},
  journal={arXiv preprint arXiv:2211.01910},
  year={2022}
}

@article{ye2023prompt,
  title={Prompt Engineering a Prompt Engineer},
  author={Ye, Qinyuan and Axmed, Maxamed and Pryzant, Reid and Khani, Fereshte},
  journal={arXiv preprint arXiv:2311.05661},
  year={2023}
}

@inproceedings{narayanan2019pipedream,
  title={PipeDream: Generalized pipeline parallelism for DNN training},
  author={Narayanan, Deepak and Harlap, Aaron and Phanishayee, Amar and Seshadri, Vivek and Devanur, Nikhil R and Ganger, Gregory R and Gibbons, Phillip B and Zaharia, Matei},
  booktitle={Proceedings of the 27th ACM Symposium on Operating Systems Principles},
  pages={1--15},
  year={2019}
}

@article{huang2019gpipe,
  title={Gpipe: Efficient training of giant neural networks using pipeline parallelism},
  author={Huang, Yanping and Cheng, Youlong and Bapna, Ankur and Firat, Orhan and Chen, Dehao and Chen, Mia and Lee, HyoukJoong and Ngiam, Jiquan and Le, Quoc V and Wu, Yonghui and others},
  journal={Advances in neural information processing systems},
  volume={32},
  year={2019}
}

@inproceedings{yu2022orca,
  title={Orca: A distributed serving system for $\{$Transformer-Based$\}$ generative models},
  author={Yu, Gyeong-In and Jeong, Joo Seong and Kim, Geon-Woo and Kim, Soojeong and Chun, Byung-Gon},
  booktitle={16th USENIX Symposium on Operating Systems Design and Implementation (OSDI 22)},
  pages={521--538},
  year={2022}
}

@article{S-LORA,
  title={S-LoRA: Serving Thousands of Concurrent LoRA Adapters},
  author={Sheng, Ying and Cao, Shiyi and Li, Dacheng and Hooper, Coleman and Lee, Nicholas and Yang, Shuo and Chou, Christopher and Zhu, Banghua and Zheng, Lianmin and Keutzer, Kurt and others},
  journal={arXiv preprint arXiv:2311.03285},
  year={2023}
}

@inproceedings{kwon2023efficient,
  title={Efficient memory management for large language model serving with pagedattention},
  author={Kwon, Woosuk and Li, Zhuohan and Zhuang, Siyuan and Sheng, Ying and Zheng, Lianmin and Yu, Cody Hao and Gonzalez, Joseph and Zhang, Hao and Stoica, Ion},
  booktitle={Proceedings of the 29th Symposium on Operating Systems Principles},
  pages={611--626},
  year={2023}
}

@misc{lightllm,
  title = {LightLLM},
  author = {{ModelTC}},
  howpublished = {\url{https://github.com/ModelTC/lightllm/}},
  year = {2025},
}

@article{chen2023punica,
  title={Punica: Multi-Tenant LoRA Serving},
  author={Chen, Lequn and Ye, Zihao and Wu, Yongji and Zhuo, Danyang and Ceze, Luis and Krishnamurthy, Arvind},
  journal={arXiv preprint arXiv:2310.18547},
  year={2023}
}

@article{liu2021p,
  title={P-tuning v2: Prompt tuning can be comparable to fine-tuning universally across scales and tasks},
  author={Liu, Xiao and Ji, Kaixuan and Fu, Yicheng and Tam, Weng Lam and Du, Zhengxiao and Yang, Zhilin and Tang, Jie},
  journal={arXiv preprint arXiv:2110.07602},
  year={2021}
}

@article{liu2023gptunderstand,
title = {GPT understands, too},
journal = {AI Open},
year = {2023},
issn = {2666-6510},
doi = {https://doi.org/10.1016/j.aiopen.2023.08.012},
url = {https://www.sciencedirect.com/science/article/pii/S2666651023000141},
author = {Xiao Liu and Yanan Zheng and Zhengxiao Du and Ming Ding and Yujie Qian and Zhilin Yang and Jie Tang},
}

@article{chen2023autoagents,
  title={AutoAgents: A Framework for Automatic Agent Generation},
  author={Chen, Guangyao and Dong, Siwei and Shu, Yu and Zhang, Ge and Sesay, Jaward and Karlsson, B{\"o}rje F and Fu, Jie and Shi, Yemin},
  journal={arXiv preprint arXiv:2309.17288},
  year={2023}
}

@inproceedings{thorpe2023bamboo,
  title={Bamboo: Making Preemptible Instances Resilient for Affordable Training of Large $\{$DNNs$\}$},
  author={Thorpe, John and Zhao, Pengzhan and Eyolfson, Jonathan and Qiao, Yifan and Jia, Zhihao and Zhang, Minjia and Netravali, Ravi and Xu, Guoqing Harry},
  booktitle={20th USENIX Symposium on Networked Systems Design and Implementation (NSDI 23)},
  pages={497--513},
  year={2023}
}

@inproceedings{jang2023oobleck,
  title={Oobleck: Resilient Distributed Training of Large Models Using Pipeline Templates},
  author={Jang, Insu and Yang, Zhenning and Zhang, Zhen and Jin, Xin and Chowdhury, Mosharaf},
  booktitle={Proceedings of the 29th Symposium on Operating Systems Principles},
  pages={382--395},
  year={2023}
}

@misc{AWS,
  title = {{Amazon Web Services}},
  howpublished = {\url{https://aws.amazon.com/}},
  note = {2025}
}

@misc{AlibabaCloud,
  title = {{Alibaba Cloud}},
  howpublished = {\url{https://www.alibabacloud.com/}},
  note = {2023}
}

@misc{Azure,
  title = {{Azure Cloud}},
  howpublished = {\url{https://azure.microsoft.com/}},
  note = {2025}
}

@Misc{claude,
howpublished = {\url{https://claude.ai/}},
note = {Accessed: 2025-01}
}

@Misc{promptbase,
howpublished = {\url{https://promptbase.com/}},
note = {Accessed: 2025-01}
}

@article{UniSched,
  title={UniSched: A Unified Scheduler for Deep Learning Training Jobs with Different User Demands},
  author={Gao, Wei and Ye, Zhisheng and Sun, Peng and Zhang, Tianwei and Wen, Yonggang},
  journal={IEEE Transactions on Computers},
  year={2024},
  publisher={IEEE}
}

@article{yang2023hydra,
  title={Hydra: Deadline-aware and Efficiency-oriented Scheduling for Deep Learning Jobs on Heterogeneous GPUs},
  author={Yang, Zichao and Wu, Heng and Xu, Yuanjia and Wu, Yuewen and Zhong, Hua and Zhang, Wenbo},
  journal={{IEEE} Trans. Computers},
  year={2023},
  publisher={IEEE}
}

@misc{textcortex,
  author       = {{TextCortex}},
  title        = {AI Prompt Marketplace - TextCortex},
  howpublished = {\url{https://textcortex.com/templates/ai-prompt-marketplace}},
  year         = {2024},
  note         = {Accessed: 2025-01}
}

@misc{promptrr,
  author       = {{Promptrr}},
  title        = {Promptrr},
  howpublished = {\url{https://promptrr.io/}},
  year         = {2024},
  note         = {Accessed: 2025-01}
}

@misc{flowgpt,
  author       = {{FlowGPT}},
  title        = {FlowGPT: Prompt Engineering for the Future},
  howpublished = {\url{https://flowgpt.com/}},
  year         = {2024},
  note         = {Accessed: 2025-01}
}

@misc{InterVenor,
      title={INTERVENOR: Prompting the Coding Ability of Large Language Models with the Interactive Chain of Repair}, 
      author={Hanbin Wang and Zhenghao Liu and Shuo Wang and Ganqu Cui and Ning Ding and Zhiyuan Liu and Ge Yu},
      year={2024},
      eprint={2311.09868},
      archivePrefix={arXiv},
      primaryClass={cs.SE},
      url={https://arxiv.org/abs/2311.09868}, 
}

@misc{PoT,
      title={Program of Thoughts Prompting: Disentangling Computation from Reasoning for Numerical Reasoning Tasks}, 
      author={Wenhu Chen and Xueguang Ma and Xinyi Wang and William W. Cohen},
      year={2023},
      eprint={2211.12588},
      archivePrefix={arXiv},
      primaryClass={cs.CL},
      url={https://arxiv.org/abs/2211.12588}, 
}

@inproceedings{RelIndex,
    title = "Large Language Models as Financial Data Annotators: A Study on Effectiveness and Efficiency",
    author = "Aguda, Toyin D.  and
      Siddagangappa, Suchetha  and
      Kochkina, Elena  and
      Kaur, Simerjot  and
      Wang, Dongsheng  and
      Smiley, Charese",
    editor = "Calzolari, Nicoletta  and
      Kan, Min-Yen  and
      Hoste, Veronique  and
      Lenci, Alessandro  and
      Sakti, Sakriani  and
      Xue, Nianwen",
    booktitle = "Proceedings of the 2024 Joint International Conference on Computational Linguistics, Language Resources and Evaluation (LREC-COLING 2024)",
    month = may,
    year = "2024",
    address = "Torino, Italia",
    publisher = "ELRA and ICCL",
    pages = "10124--10145",
}

@inproceedings{ymir,
author = {Gao, Wei and Zhuang, Weiming and Li, Minghao and Sun, Peng and Wen, Yonggang and Zhang, Tianwei},
title = {Ymir: A Scheduler for Foundation Model Fine-tuning Workloads in Datacenters},
year = {2024},
address = {New York, NY, USA},
booktitle = {Proceedings of the 38th ACM International Conference on Supercomputing},
pages = {259–271},
}

@inproceedings{autosched,
author = {Gao, Wei and Zhang, Xu and Huang, Shan and Guo, Shangwei and Sun, Peng and Wen, Yonggang and Zhang, Tianwei},
title = {AutoSched: An Adaptive Self-configured Framework for Scheduling Deep Learning Training Workloads},
year = {2024},
isbn = {9798400706103},
publisher = {Association for Computing Machinery},
address = {New York, NY, USA},
pages = {473–484},
numpages = {12},
location = {Kyoto, Japan},
series = {ICS '24}
}

@misc{nvidia2020a100whitepaper,
	title        = {NVIDIA A100},
	year         = 2022,
	howpublished = {\url{https://www.nvidia.com/en-sg/data-center/a100/}}
}

@misc{nvidia2022h100whitepaper,
	title        = {NVIDIA H100},
	year         = 2022,
	howpublished = {\url{https://resources.nvidia.com/en-us-tensor-core/gtc22-whitepaper-hopper}}
}

@misc{claude_ai,
  author       = {Anthropic},
  title        = {Claude AI},
  howpublished = {\url{https://claude.ai/}},
  year         = {2025}
}

@misc{google_gemini,
  author       = {Google},
  title        = {Gemini AI},
  howpublished = {\url{https://gemini.google.com/}},
  year         = {2025}
}

@misc{openai2023gpt4,
  author       = {OpenAI},
  title        = {ChatGPT},
  howpublished = {\url{https://chatgpt.com/}},
  year         = {2025}
}

@article{honovich2022instruction,
  title={Instruction induction: From few examples to natural language task descriptions},
  author={Honovich, Or and Shaham, Uri and Bowman, Samuel R and Levy, Omer},
  journal={arXiv preprint arXiv:2205.10782},
  year={2022}
}

@misc{portkeyai,
  author       = {Portkey AI},
  title        = {Portkey AI},
  year         = {2025},
  url          = {https://portkey.ai/features/prompt-management},
  note         = {Accessed: 2025-01}
}

@article{wang2025sequential,
  author    = {Shuyang Wang and Somayeh Moazeni and Diego Klabjan},
  title     = {A Sequential Optimal Learning Approach to Automated Prompt Engineering in Large Language Models},
  journal   = {arXiv preprint arXiv:2501.03508v1},
  year      = {2025},
  url       = {https://arxiv.org/abs/2501.03508v1},
  month     = {January},
  day       = {7}
}

@article{do2024automatic,
  title={Automatic Prompt Selection for Large Language Models},
  author={Do, Viet-Tung and Hoang, Van-Khanh and Nguyen, Duy-Hung and Sabahi, Shahab and Yang, Jeff and Hotta, Hajime and Nguyen, Minh-Tien and Le, Hung},
  journal={arXiv preprint arXiv:2404.02717},
  year={2024}
}

@inproceedings{fu2024serverlessllm,
  title={ServerlessLLM: Low-latency serverless inference for large language models},
  author={Fu, Yao and Xue, Leyang and Huang, Yeqi and Brabete, Andrei-Octavian and Ustiugov, Dmitrii and Patel, Yuvraj and Mai, Luo},
  booktitle={18th USENIX Symposium on Operating Systems Design and Implementation},
  pages={135--153},
  year={2024},
  organization={USENIX Association}
}

@misc{manus_website,
  title        = {Manus},
  author       = {{Manus}},
  howpublished = {\url{https://manus.im/}},
  year         = {2025-04}, 
}

@article{cobbe2021gsm8k,
  title={Training Verifiers to Solve Math Word Problems},
  author={Cobbe, Karl and Kosaraju, Vineet and Bavarian, Mohammad and Chen, Mark and Jun, Heewoo and Kaiser, Lukasz and Plappert, Matthias and Tworek, Jerry and Hilton, Jacob and Nakano, Reiichiro and Hesse, Christopher and Schulman, John},
  journal={arXiv preprint arXiv:2110.14168},
  year={2021}
}

@article{chen2025sppo,
  author  = {Chen, Qiaoling and Li, Shenggui and Gao, Wei and Sun, Peng and Wen, Yonggang and Zhang, Tianwei},
  title   = {SPPO: Efficient Long-sequence LLM Training via Adaptive Sequence Pipeline Parallel Offloading},
  journal = {arXiv preprint arXiv:2503.10377},
  year    = {2025},
  month   = mar,
  day     = {13},
  eprint  = {2503.10377},
  archivePrefix = {arXiv},
  primaryClass  = {cs.LG}
}

@article{gao2025icefrog,
  author  = {Gao, Wei and Ouyang, Zhuoyuan and Sun, Peng and Zhang, Tianwei and Wen, Yonggang},
  title   = {ICEFROG: A Layer-Elastic Scheduling System for Deep Learning Training in GPU Clusters},
  journal = {IEEE Transactions on Parallel and Distributed Systems},
  year    = {2025},
  volume  = {36},
  number  = {6},
  month   = jun
}

@inproceedings{gao2025rethinkingkvcache,
  author    = {Gao, Wei and Zhou, Xinyu and Sun, Peng and Zhang, Tianwei and Wen, Yonggang},
  title     = {Rethinking Key-Value Cache Compression Techniques for Large Language Model Serving},
  booktitle = {Proceedings of the Annual Conference on Machine Learning and Systems (MLSys)},
  year      = {2025},
  month     = may
}

@article{guo2025adaptis,
  author        = {Guo, Jihu and Ma, Tenghui and Gao, Wei and Sun, Peng and Li, Jiaxing and Chen, Xun and Jin, Yuyang and Lin, Dahua},
  title         = {AdaPtis: Reducing Pipeline Bubbles with Adaptive Pipeline Parallelism on Heterogeneous Models},
  journal       = {arXiv preprint arXiv:2509.23722},
  year          = {2025},
  month         = sep,
  day           = {28},
  eprint        = {2509.23722},
  archivePrefix = {arXiv},
  primaryClass  = {cs.LG}
}



\end{document}